\documentstyle[11pt]{article}
\newcommand{\mysection}[1]{\section{#1}\setcounter{equation}{0}}

\def\bea{\begin{eqnarray}} 
\def\eea{\end{eqnarray}}
\def\beann{\begin{eqnarray*}} 
\def\eeann{\end{eqnarray*}}
\def\beq{\begin{equation}} 
\def\eeq{\end{equation}}
\def\ba{\begin{array}} 
\def\ea{\end{array}}
\def\ben{\begin{enumerate}} 
\def\een{\end{enumerate}}

\def\5{\bar }  
\def\6{\partial } 
\def\7{\hat } 
\def\4{\tilde }

\def\gh{\mbox{gh}} 
\def\agh{\mbox{antigh}}
\def\tot{\mbox{totdeg}}
\def\deg{\mbox{formdeg}}

\def\cA{{\cal A}}
\def\cB{{\cal B}}
\def\cD{{\cal D}} 
\def\cF{{\cal F}}
\def\cL{{\cal L}}
\def\cR{{\cal R}} 
\def\cT{{\cal T}} 
\def\cU{{\cal U}} 
\def\cV{{\cal V}}
\def\cW{{\cal W}}
\def\cZ{{\cal Z}}

\def\TIX{{\, \imath}} 
\def\TJX{\jmath}
\def\Indx{{\7r}} 
\def\Jndx{{\7s}}

\def\s0#1#2{\mbox{\small{$\frac{#1}{#2}$}}}
\def\chris#1#2#3{{\Gamma_{#1#2}}^{#3}}
\def\f#1#2#3{{f_{#1#2}}^{#3}}
\def\A#1#2{{A_{#2}}^{#1}}
\def\e#1#2{v_{#2}{}^{#1}} 
\def\E#1#2{V_{#1}{}^{#2}}
\def\AA #1#2{{{\cal A}_{#2}}^{#1}}
\def\viel#1#2{e_{#2}{}^{#1}} 
\def\Viel#1#2{E_{#1}{}^{#2}}
\def\csum#1#2{\sum_{#1}\hspace{-1.#2em}\circ\ \ \ }

\begin{document}
\begin{titlepage}

\begin{flushright}
KUL--TF--96/8\\
hep-th/9604025\\
to appear in {\em Commun. Math. Phys.}
\end{flushright}
\vfill

\begin{center}
{\LARGE Local BRST Cohomology and Covariance}
\end{center}
\vfill

\begin{center}
{\large
Friedemann Brandt\,$^*$}
\end{center}
\vfill

\begin{center}{\sl
 Instituut voor Theoretische Fysica, Katholieke
 Universiteit Leuven,\\
 Celestijnenlaan 200 D, B--3001 Leuven, Belgium
}\end{center}
\vfill

\begin{abstract}
The paper provides a framework for a systematic analysis of the local 
BRST cohomology in a large class of gauge theories. The approach is 
based on the cohomology of $s+d$ in the jet space of fields and 
antifields, $s$ and $d$ being the BRST operator and exterior derivative
respectively. It relates the BRST cohomology to an underlying 
gauge covariant algebra and reduces its computation to a compactly 
formulated problem involving only suitably defined generalized 
connections and tensor fields.
The latter are shown to provide the building blocks of physically
relevant quantities such as gauge invariant actions, Noether currents
and gauge anomalies, as well as of the equations of motion.
\end{abstract}

\vspace{10em}

\hrule width 5.cm
\vspace*{.5em}

{\small \noindent $^*$ Junior fellow of the research council (DOC)
of the K.U. Leuven.}
\end{titlepage}


\mysection{Introduction}\label{intro}

\subsection{Motivation}
\label{intro1}

Gauge invariance underlies as a basic principle our present
models of fundamental interactions and is widely
used when one looks for extensions of these models.
The BRST-BV formalism provides a general framework
to deal with many aspects of gauge symmetry,
both in classical and quantum field theory.
It was first established by Becchi, Rouet and
Stora  \cite{brs} in the
context of renormalization of abelian Higgs--Kibble and
Yang--Mills gauge theories,
later extended by Kallosh to supergravity with open
gauge algebra \cite{K} (see also \cite{STvN})
and by de Wit and van Holten
to general gauge theories \cite{dWvH},
resulting finally in the universal field-antifield formalism
of Batalin and Vilkovisky \cite{bv} which allows to treat
all kinds of gauge theories within an elegant unified framework.
The usefulness of this formalism is
mainly based on the fact that it encodes the
gauge symmetry and all its properties
in a single antiderivation which is strictly nilpotent
on all the fields and antifields. Throughout this paper, this
antiderivation is called the BRST operator
and denoted by $s$.

The nilpotency of $s$ establishes in particular
the local BRST cohomology, i.e. the cohomology of $s$ in the 
space of local functionals (= integrated local volume forms)
of the fields and antifields.
This cohomology has many physically relevant applications.
It determines for instance
gauge invariant actions and their consistent
deformations \cite{bh}, the dynamical local
conservation laws \cite{bbh1} and the possible gauge anomalies
(see e.g.\ \cite{wz,brs,StoraZumino,TvNvP,piguetsorella,gp}) 
of a gauge theory and
is a useful tool in the renormalization
of quantum field theories even when a theory is not
renormalizable in the usual sense \cite{gw}.

Since the BRST cohomology can be defined
for any gauge theory and since the
correspondence of its cohomology classes to the
mentioned physical quantities is universal too,
it is worthwhile to look for a suitable general framework within
which this cohomology can be computed efficiently and
which has a large range of applicability.

The purpose of this paper is to propose such a framework.
It applies to a large class of gauge theories and
relates the BRST cohomology to an underlying
gauge covariant algebra. This includes a definition
of tensor fields on which this algebra is realized and
of generalized connections associated with it, and reduces
the computation of the BRST cohomology locally to
a problem involving only these quantities.

The reduced problem is formulated
very compactly in terms of identities analogous to 
the ``Russian formula" in Yang--Mills theory 
\cite{StoraZumino,baumieg}%
\footnote{Orinally the term ``Russian formula"
was introduced by Stora in the second
ref.\ in \cite{StoraZumino} for a 
different but related identity. Here it is
used as in the last ref.\ in \cite{StoraZumino}.},
\beq 
F=(s+d)(C+A)+(C+A)^2.
\label{i1}
\eeq
Here $C$, $A$ and $F$ are the familiar Lie algebra 
valued Yang--Mills
ghost fields, connection and curvature forms respectively, 
$s$ is the 
Yang--Mills BRST operator, and $d$ is the spacetime
exterior derivative. The usefulness of
(\ref{i1}) is based, among others, on its remarkable
property to compress the familiar BRST transformations
of the Yang--Mills ghost and gauge fields, as well as the
construction of the field strength 
in terms of the gauge field
into a single identity. The combination $C+A$ occurring
in (\ref{i1}) is an example of what will be called 
a generalized connection here.

\subsection{Relations and differences to other approaches}
\label{other}

The proposed approach generalizes a concept
outlined in \cite{ten} (see also \cite{dragon})
for the study of the ``restricted" (= antifield independent) 
BRST cohomology in a special class of gauge theories characterized
among others by
(a) the presence of (spacetime) diffeomorphisms among the gauge
symmetries, (b) the closure and irreducibility of the gauge
algebra, (c) the presence of `enough' independent gauge fields
ensuring that all the derivatives of the ghost 
fields can be eliminated
from the BRST cohomology. In such theories, the extension of the
concept of \cite{ten} to the full cohomological problem, 
including the
antifields, is (more or less) straightforward
and was used already in
\cite{rapid,bbhgrav} within a complete computation of the
BRST cohomology in 
Einstein gravity and Einstein--Yang--Mills theories.

Here these ideas are
extended to general gauge theories. In particular none
of the conditions (a)--(c) is needed as a prerequisite for the
methods outlined in this paper. This is possible thanks to
suitable generalizations of the concept \cite{ten}
which at the same time modify and unite various 
techniques that have been developed over the last 20 years, thereby
revealing relations between them which are less apparent
in other approaches. Such techniques, to be
described later in detail, 
are the so-called descent equation technique,
the use of contracting homotopies in jet spaces, compact
formulations of the BRST algebra analogous to the ``Russian
formula'' (\ref{i1}), and spectral sequence techniques 
along the lines of homological 
perturbation theory \cite{HPT1,HPT2,henfisch}.
Let me now briefly comment on the use
of these techniques in this paper, as compared to
other approaches.

Descent equations and the ``Russian
formula'' were first used within the celebrated
differential geometric construction of (representatives of)
chiral anomalies in
$D=2n$ dimensions from characteristic classes in $D+2$
dimensions \cite{StoraZumino}, and also within
the classification of such anomalies in \cite{talon}.
Later it became clear that the descent equations
are useful not only in connection with chiral anomalies, but
to analyse the complete BRST cohomology, cf.\ e.g.\ 
\cite{grav,hental,ten,piguetsorella,dragon}. The reason is
that they allow to deal efficiently with the total derivatives
into which the integrands of BRST invariant
functionals transform in general.

In this paper we will compress the descent equations into a 
compact form. To this end the BRST operator $s$ and 
the exterior derivative $d$ will be united
to the single operator
\beq 
\4s=s+d
\label{des4}
\eeq
defined on local ``total forms" (see section \ref{setting}).
This idea is not new; in fact it is familiar
from the construction and classification of chiral 
anomalies mentioned above. However, somewhat surprisingly, it was
not utilized systematically in a general approach to the 
BRST cohomology on local functionals later.

The systematic use of $\4s$ is fundamental 
to the method proposed here and has several advantages. 
In particular it
allows us to extend the concept of \cite{ten}
to theories which do not satisfy the assumptions (a)--(c)
mentioned above, such as Yang--Mills theory whose BRST 
cohomology has been calculated by different means in
\cite{com,hental,bhym,bbh2}. The use of $\4s$
is particularly well adapted to the analysis of the
BRST cohomology on local functionals
because the latter is in fact isomorphic
to the cohomology of $\4s$ on local total forms, 
at least locally, c.f.\ \cite{dragon} 
and section \ref{descent}%
\footnote{The isomorphism applies only
to the BRST cohomology on local functionals, i.e.\ to the
relative cohomology $H(s|d)$ on local {\em volume} forms. 
It does not extend to $H(s|d)$ 
at lower form degrees in general.}.

Contracting homotopies similar to the ones 
used here were constructed and applied to BRST 
cohomological problems e.g.\ already in 
\cite{com,grav,hental}. However, these contracting homotopies
were designed for the cohomology of $s$ \cite{hental} and its
linearized version \cite{com,grav} respectively.
The method proposed here extends them 
to the $\4s$-cohomology. This has the important consequence
that it leads directly to the mentioned
compact formulation of the cohomological problem
in terms of identities analogous to the
``Russian formula'' (\ref{i1}).
For instance, when applied to Yang--Mills theory, the 
contracting homotopy for $\4s$
singles out the special combination (generalized
connection) $A+C$ occurring in (\ref{i1}). As a result,
(\ref{i1}) itself arises naturally in this approach,
cf.\ section \ref{ym}. In contrast, the corresponding
contracting homotopy \cite{hental} for $s$ gives instead of $A+C$ 
just $C$ and makes no contact 
with the ``Russian formula'' (it does however provide
the same tensor fields).

The proposed approach also extends
the methods developed in \cite{henfisch} to use and
deal with the antifields along the lines of
homological perturbation theory \cite{HPT1,HPT2}. 
This extension is straightforward and, again,
related to the use of
$\4s$ instead of $s$. Among others it will allow us to
trace the BRST cohomology at all
ghost numbers (including negative ones) back to a
weak (= on-shell) cohomological problem 
involving the tensor fields and generalized connections only.
This has been utilized recently in \cite{SUGRA} in order to
compute the BRST cohomology in four dimensional $N=1$
supergravity.

Finally, the approach provides a `cohomological' perspective on
tensor fields and connections. The latter
are usually characterized through
specific transformation properties 
under the respective symmetries.
However, in a general gauge theory it is not always clear
from the outset  which transformation laws should be imposed
for this purpose. An advantage of the
approach proposed here is that such transformation laws
need not be specified from the start. Rather, they
emerge from the approach itself. Such a
characterization of tensor fields,
connections and the corresponding
transformation laws has two major advantages: (i)
it is purely algebraic and does not
invoke any concepts in addition to the BRST cohomology
itself; (ii) it is physically meaningful
because the resulting tensor fields
and generalized connections
turn out to provide among others
the building blocks of gauge invariant actions, Noether currents,
anomalies and of the equations of motions.

\subsection{Outline of the paper}
\label{outline}

The paper has been organized as follows. 
Section \ref{setting}
sketches the basic algebraic approach to the BRST cohomology
used in this paper and introduces some terminology and
notation. Sections \ref{descent}
and \ref{red} relate the local BRST cohomology
to the cohomology of  $\4s$ and its weak (= ``on-shell")
counterpart. Section \ref{trivpairs} introduces
the concept of contracting homotopies for $\4s$ in jet spaces, and
section \ref{ten} shows that this concept is intimately
related to the existence of a gauge covariant algebra and
a compact formulation of the BRST algebra on
tensor fields and generalized connections.
Section \ref{ex} illustrates the method for various examples
which do not satisfy the aforementioned assumptions
(a)--(c) of \cite{ten}
(the examples are Yang--Mills theory, 
Einstein gravity in the metric formulation,
supergravity with open gauge algebra and two-dimensional Weyl
invariant sigma models). Sections
\ref{structure}--\ref{EOM} spell out implications
for the structure of gauge invariant actions,
Noether currents, gauge anomalies, etc., as well as for the
classical equations of motion. In section \ref{xdep} a
special aspect of the cohomological problem is discussed,
concerning the
explicit dependence of the solutions on the 
coordinates of the base manifold which will be called
``spacetime'' henceforth, for no reason at all. 
The paper is ended by some 
concluding remarks in
section \ref{conclusion} and two appendices
containing details concerning
the algebraic approach and conventions used in the paper.

\mysection{Algebraic setting, definitions and notation}
\label{setting}

In order to define the local BRST cohomology in a particular
theory one has to specify the BRST operator $s$ and the space
in which its cohomology is to be computed. The BRST operator
is defined on a
set of fields $\Phi^A$ and corresponding antifields
$\Phi^*_A$ according to standard rules of the 
field-antifield formalism
summarized in appendix \ref{brs}. In particular these rules
include that the BRST operator is nilpotent and commutes with the
spacetime derivatives $\6_\mu$,
\beq s^2=s\6_\mu-\6_\mu s=\6_\mu\6_\nu-\6_\nu\6_\mu=0.
\label{des1}\eeq
The basic concept underlying these
fundamental relations and the whole paper
is the {\em jet bundle approach} \cite{jet}
sketched in the appendix \ref{jet}.
Essentially this means simply
that the fields, antifields and all their derivatives
are understood as
local coordinates of an infinite jet space. For this set of
jet coordinates the
collective notation $[\Phi,\Phi^*]$ is used.
The local jet coordinates are completed by 
the spacetime coordinates
$x^\mu$ and the differentials $dx^\mu$. The differentials are
counted among the jet coordinates by pure convention 
and convenience.
The derivatives $\6_\mu$ are
defined as total derivative operators in the jet space,
cf.\ eq.\ (\ref{totder}),
and become usual partial derivatives
on the local sections of the jet bundle.

The concrete
BRST transformations of the fields and antifields depend on the
particular theory and its gauge symmetry,
whereas the spacetime coordinates
$x^\mu$ and differentials $dx^\mu$
are always BRST invariant in accordance with the
second relation (\ref{des1}),
\beq s\, x^\mu=0,\quad s\, dx^\mu=0.\label{des2}\eeq
The use of the differentials is in principle not necessary
but turns out to be very useful
in order to analyse the local BRST cohomology.
In particular it allows to
define $d=dx^\mu\6_\mu$ and $\4s=s+d$ in the jet space.
The relations (\ref{des1}) are equivalent to the 
nilpotency of $\4s$,
\beq \4s^2=0\quad\Leftrightarrow\quad s^2=sd+ds=d^2=0.
\label{des5}\eeq

The usefulness of $\4s$ in the context of the local BRST cohomology
stems from the fact that it allows to write and analyse
the descent equations in a compact form (cf.\ section
\ref{descent}). The descent
equations involve {\em local $p$-forms}
\beq \omega_p=\frac 1{p!}\, dx^{\mu_1}\ldots dx^{\mu_p}
\omega_{\mu_1\ldots\mu_p}(x,[\Phi,\Phi^*]).
\label{des6}\eeq
These forms are required to be local
in the sense that they are formal series'
in the antifields, ghosts and their derivatives 
such that each piece
with definite antighost number (cf.\ \cite{henfisch} and section
\ref{red}) depends polynomially
on the derivatives of all the fields and antifields.
{}From the outset no
additional requirements are imposed on local forms here.
In particular they are not restricted by power counting,
it is not assumed that the
indices $\mu_i$ of the functions $\omega_{\mu_1\ldots\mu_p}$ occurring
in (\ref{des6}) indicate their actual transformation
properties under Lorentz or general coordinate transformations, and
local forms are not required to be globally well-defined
in whatever sense.

A {\em local functional} is by definition an integrated 
local volume
form $\int \omega_D$ (throughout this paper $D$ denotes the
spacetime dimension). It is called BRST invariant if
$s\omega_D$ is $d$-exact in the space of local forms,
i.e.\ if $s\omega_D+d\omega_{D-1}=0$ holds for some local form
$\omega_{D-1}$. Translated to the local sections of the
jet bundle, in general this requires local functionals
to be BRST invariant only up to surface integrals.
Analogously a local functional $\int \omega_D$ is called 
BRST-exact (or trivial)
if $\omega_D=s\eta_D+d\eta_{D-1}$ holds for some local forms
$\eta_D$ and $\eta_{D-1}$.
The BRST cohomology on local functionals considered here is thus
actually the relative cohomology $H(s|d)$ 
of $s$ and $d$ on local volume forms.
This cohomology
is well-defined due to (\ref{des5}) and represented by
solutions $\omega_D$ of
\beq
s\omega_D+d\omega_{D-1}=0,\quad
\omega_D\neq s\eta_D+d\eta_{D-1}\ .
\label{des8b}\eeq

In the next section $H(s|d)$ will be related
to the cohomology of $\4s$ on
local {\em total forms} $\4\omega$. The latter are by definition
formal sums of local forms with different form degrees,
\beq \4\omega=\sum_p\omega_p\ .\label{des11}\eeq
The $\4s$-cohomology on local total forms is then defined
through the condition $\4s\4\omega=0$ modulo trivial solutions
of the form $\4s\4\eta+constant$ where $\4\eta$ 
is a local total form
and the constant is included for convenience. The
representatives of this cohomology are thus
local total forms $\4\omega$ solving
\beq \4s\4\omega=0,\quad \4\omega\neq\4s\4\eta+constant.
\label{des12}\eeq
The natural degree in the space of local total forms is
the sum of the ghost number ($\gh$) and the form degree ($\deg$),
called the {\em total degree} ($\tot$),
\beq \tot=\gh+\deg\ .\label{totdef}\eeq
A local total form with
definite total degree $G$ is thus a sum of local
$p$-forms with ghost number $g=G-p$ ($p=0,\ldots,D$).
$\4s$ has total degree 1, i.e.\ it maps
a local total form with total degree $G$ to another
one with total degree $G+1$.

\mysection{Descent equations}\label{descent}

It is easy to see that the BRST cohomology on local
functionals is locally isomorphic
to the cohomology of $\4s$ on local total forms%
\footnote{Here and in the following
local equalities or isomorphisms refer to sufficiently 
small patches
of the jet space. Global properties of the jet bundle are not taken
into account.}.
To show this, one only needs (\ref{des5}) and a theorem on
the cohomology of $d$ on local forms, sometimes called the
{\em algebraic Poincar\'e lemma}. The latter
states that locally  any $d$-closed local
$p$-form is $d$-exact for $0<p<D$ and constant for $p=0$, 
while local
volume forms ($p=D$) are locally $d$-exact if and only if they have
vanishing Euler--Lagrange derivative with respect to all the fields
and antifields \cite{various,com,dragon}.

The local isomorphism of the cohomological problems 
associated with (\ref{des8b}) and (\ref{des12}) can be derived
by standard arguments which are therefore only sketched. Let me
therefore sketch its derivation. The arguments are
standard and therefore not given in detail. Suppose that
$\omega_D$ solves $s\omega_D+d\omega_{D-1}=0$.
Applying $s$ to this equation results in
$d(s\omega_{D-1})=0$ due to (\ref{des5}). 
Hence, $s\omega_{D-1}$ is
$d$-closed. Since it is not a volume form, it is thus
also $d$-exact in the space of local forms 
according to the algebraic
Poincar\'e lemma. Hence, there is a (possibly vanishing) local
$(D-2)$-form $\omega_{D-2}$ satisfying 
$s\omega_{D-1}+d\omega_{D-2}=0$.
Iterating the arguments
one concludes the existence of a set of local forms $\omega_{p}$,
$p=p_0,\ldots,D$ satisfying
\beq  s\omega_{p}+d\omega_{p-1}=0 \quad \mbox{for }D\geq p>p_0;
\quad s\omega_{p_0}=0 \label{des9}\eeq
for some $p_0$.
These equations are called the descent equations\footnote{For
$p_0=0$ the algebraic Poincar\'e lemma alone actually implies only
$s\omega_0=const.$; however, in meaningful gauge theories
a BRST-exact constant vanishes necessarily, as one
easily verifies (note that a constant can occur only 
if $\omega_0$ has ghost number $-1$). 
Notice that this might not hold
anymore if one extends the BRST--BV formalism by
including constant ghosts corresponding
e.g.\ to global symmetries \cite{ten,dragon}. Such an extension 
is always possible \cite{bhw}
but not considered here.}.
They can be compactly written in the form
\[ \4s\, \4\omega=0,\quad \4\omega=\sum_{p=p_0}^D\omega_{p}\ .\]
Hence, any solution of $s\omega_D+d\omega_{D-1}=0$ corresponds to
an $\4s$-closed local total form and the reverse is evidently also
true. Using again the algebraic Poincar\'e lemma and (\ref{des5}),
it is easy to see that $\omega_D$ is a trivial solution
of the form $s\eta_{D}+d\eta_{D-1}$ if and only if
$\4\omega$ is trivial too, i.e.\ if and only if
$\4\omega=\4s\4\eta+constant$. Since $\4\omega$ has total
degree $(g+D)$ if $\omega_D$ has ghost number $g$ we conclude
\medskip

\noindent {\bf Lemma 3.1:}
{\em The BRST-cohomology on local functionals with ghost number $g$
and the $\4s$-cohomology on local total forms
of total degree $G=g+D$ are locally isomorphic. That is
to say, locally the solutions of (\ref{des8b}) with
ghost number $g$ correspond one-to-one (modulo trivial solutions)
to the solutions of (\ref{des12}) with total degree $G=g+D$.}

\mysection{Equivalence to the weak
cohomology of $\4\gamma=\gamma+d$}\label{red}

A simple and useful concept in the study of the BRST cohomology is
a suitable expansion of local functionals and forms in
powers of the antifields.
Following the lines of \cite{henfisch} it will now be used
to show that the $\4s$-cohomology on local total
forms of the fields and antifields reduces to a weak (= on-shell)
cohomology on antifield independent local total forms.

The most useful expansion in the antifields takes their respective
ghost numbers into account. This is achieved
through the so-called
antighost number ($\agh$) defined according to
\beq \agh (\Phi^*_A)=-\gh(\Phi^*_A),\quad
\agh (\Phi^A)=\agh (dx^\mu)=\agh (x^\mu)=0.\label{agh}\eeq
In particular the BRST operator can be decomposed
into pieces with definite antighost number (one says a piece
has antighost number $k$ if
it raises the antighost number by $k$ units).
The decomposition
of $s$ starts always with a piece of antighost number $-1$,
\beq s=\delta+\gamma+\sum_{k\geq 1}s_k\ ,\quad
\agh(\delta)=-1,\quad \agh(\gamma)=0,\quad
\agh (s_k)=k.\label{sdec}\eeq
The most important pieces in this decomposition are $\delta$ and
$\gamma$; the other pieces have positive antighost number
and play only a secondary role
in the cohomological analysis. $\delta$ is the so-called 
Koszul--Tate
differential and is nonvanishing only on the antifields,
\beq \delta\Phi^A=0,\quad
\delta\phi^*_i=\frac{\7\6^R\cL_{cl}}{\7\6\phi^i}\ ,\quad\ldots
\label{eom}\eeq
where ${\7\6^R\cL_{cl}}/{\7\6\phi^i}$ denotes the Euler--Lagrange
right-derivative of the classical Lagrangian $\cL_{cl}$ 
w.r.t. $\phi^i$.
In particular $\delta$ thus implements the classical equations of
motion in the cohomology. $\gamma$
encodes the gauge transformations because
$\gamma\phi^i$ equals a gauge transformation
of $\phi^i$ with parameters replaced by ghosts,
\beq \gamma\phi^i=R_\alpha^i C^\alpha\ ,\label{gammatrafo}\eeq
where the notation of appendix \ref{brs} is used.

(\ref{sdec}) extends straightforwardly to the
analogous decomposition of $\4s=s+d$ into pieces with
definite antighost numbers. Since $d$ has vanishing antighost
number, one simply gets
\beq \4s= \delta+\4\gamma+\sum_{k\geq 1}s_k\label{4sdec}\eeq
with
\beq \4\gamma=\gamma+d.\label{4gamma}\eeq
Note that $\4s^2=0$ decomposes into
\beq \delta^2=0,\quad
\delta\4\gamma+\4\gamma\delta=0,\quad
\4\gamma^2=-(\delta s_1+s_1\delta),\quad\ldots
\label{cms}\eeq

The usefulness of the decomposition (\ref{4sdec}) is due to the
acyclicity of the Koszul--Tate differential $\delta$ on
local functions
at positive antighost number \cite{HPT2,henfisch,VVP}. This means 
that the cohomology of $\delta$ on local total forms is
trivial at positive antighost number,\footnote{An analogous
statement does {\em not} hold for the relative cohomology of
$\delta$ and $d$. Indeed there
are in general solutions of (\ref{des8b}) which
contain no antifield independent part.
Such solutions correspond to local conservation laws \cite{bbh1}.}
\beq \delta\4\omega_k=0,\quad \agh(\4\omega_k)=k>0
\quad\Rightarrow\quad
\4\omega_k=\delta\4\eta_{k+1}\ .\label{decohom}\eeq
Using standard arguments of spectral sequence
techniques which are not repeated here, 
one concludes from
(\ref{decohom}) immediately that a nontrivial solution of
$\4s\4\omega=0$ contains necessarily an antifield independent part
$\4\omega_0$ solving
\beq \4\gamma\4\omega_0\approx 0,\quad
\4\omega_0\not\approx \4\gamma\4\eta_0+constant,\quad
\agh(\4\omega_0)=0
\label{weak7}\eeq
where $\approx$ denotes {\em weak equality} defined through
\beq A_0\approx 0
\quad:\Leftrightarrow\quad\exists A_1:\quad A_0=\delta A_1\quad 
(\agh(A_k)=k).
\label{weak}\eeq
Note that the weak equality is an ``on-shell equality"
since, due to (\ref{eom}),
$A_0\approx 0$ implies that $A_0$ vanishes for solutions of the
classical equations of motion.

Furthermore (\ref{cms}) and
(\ref{decohom}) imply that each solution
$\4\omega_0$ of (\ref{weak7}) can be completed to a 
nontrival solution
$\4\omega=\4\omega_0+\ldots$ of (\ref{des12}) and that
two different completions with the same antifield
independent part are equivalent in the
cohomology of $\4s$ (the latter follows 
immediately from the fact that
the difference of two such completions
has no antifield independent part).
This establishes the following result:
\medskip

\noindent {\bf Lemma 4.1:}
{\em The cohomology of $\4s$ on local total forms
is isomorphic to the weak cohomology of $\4\gamma$ on antifield
independent local total forms. That is to say,
any solution $\4\omega$ of (\ref{des12}) contains
an antifield independent part $\4\omega_0$ solving
(\ref{weak7}), and any solution $\4\omega_0$ of
(\ref{weak7}) can be completed
to a solution of (\ref{des12}) 
which (for fixed $\4\omega_0$) is unique
up to $\4s$-exact contributions.}
\medskip

\noindent Remark:

The weak cohomology of $\4\gamma$ on antifield
independent local total forms is well-defined since $\4\gamma$ is
weakly nilpotent on these forms, 
\beq \agh(A_0)=0\quad\Rightarrow\quad
\4\gamma^2 A_0 \approx 0.
\label{weak8}\eeq
This follows immediately from the third identity
(\ref{cms}) due to $\delta A_0=0$.

\mysection{Elimination of trivial pairs}\label{trivpairs}

A well-known technique in the study of cohomologies is the
use of contracting homotopies. I will now describe how one can
apply it within the computations of the $\4s$-cohomology
and of the weak $\4\gamma$-cohomology
introduced in the previous sections. The
idea is to construct contracting homotopy operators which
allow to eliminate
certain local jet coordinates, called {\em trivial pairs}, from
the cohomological analysis. This reduces
the cohomological problem to an analogous one involving
only the remaining jet coordinates.
For that purpose one needs to construct
suitable sets of jet coordinates
replacing the fields, antifields and their 
derivatives and satisfying
appropriate requirements. In this section I will specify such
requirements and show that they allow to eliminate
trivial pairs.
In section \ref{ex} various explicit
examples will be discussed to illustrate how one constructs
these special jet coordinates in practice.

The contracting homotopies and the trivial pairs for the 
$\4s$- and the
weak $\4\gamma$-cohomology are usually closely related.
Nevertheless,
in practical computations the use of one or the other may be
more convenient. Moreover it is often
advantageous to combine them. For instance one may first
use a contracting homotopy for the
$\4s$-cohomology that
eliminates some fields or antifields
completely, such as the antighosts and
the corresponding Nakanishi--Lautrup auxiliary fields used for
gauge fixing, and
then analyse the remaining problem by investigating
the weak $\4\gamma$-cohomology.
The arguments will be worked out in detail only for
the weak $\4\gamma$-cohomology which is
more subtle due to the occurrence of weak instead of strict
equalities. In contrast, the $\4s$-cohomology can be
treated using standard arguments which imply:
\medskip

{\bf Lemma 5.1:} {\em Suppose there is a set of
local jet coordinates
$\cB=\{\cU^\ell,\cV^\ell,\cW^i\}$ such that the change of
local jet coordinates from $\{[\Phi^A,\Phi^*_A],x^\mu,dx^\mu\}$ 
to $\cB$
is local and locally invertible%
\footnote{I.e.\ locally
any local total form $f([\Phi,\Phi^*],dx,x)$
can be uniquely expressed as a local total form $g(\cU,\cV,\cW)$ and
vice versa.} and
\bea \4s\, \cU^\ell&=&\cV^\ell\quad \forall\, \ell\ ,\label{IIA}\\
\4s\, \cW^i&=&\cR^i(\cW)\quad \forall\, i\ .\label{IIB}\eea
Then locally the $\cU$'s and $\cV$'s can be eliminated from the
$\4s$-cohomology, i.e.\ the latter reduces locally to the
$\4s$-cohomology on
local total forms depending only on the $\cW$'s.}
\medskip

The $(\cU^\ell,\cV^\ell)$ are called trivial pairs. 
As already mentioned,
lemma 5.1 can be used in particular to eliminate the antighosts,
Nakanishi--Lautrup fields and their antifields completely from the
cohomological analysis because they (and all their derivatives) 
form
trivial pairs, cf.\ e.g.\ \cite{com} and \cite{bbh1}, section 14.
In the following these
fields will be therefore neglected without loss of generality.

Let me now turn to the derivation of an analogous result for the
weak $\4\gamma$-cohomology on antifield independent 
local total forms.
Let us assume that there is a local and
locally invertible change of
jet coordinates from the antifield independent set
$\{[\Phi^A],x^\mu,dx^\mu\}$
to $\{U^\ell, V^\ell, W^i\}$ such that\footnote{One may
replace the equalities in (\ref{iia}) and (\ref{iib}) by
weak equalities without essential changes in the following
arguments.}
\bea \4\gamma U^\ell&= &V^\ell \quad \forall\, \ell\ ,\label{iia}\\
\4\gamma W^i&= &R^i(W)\quad\forall \, i\ .\label{iib}\eea
Furthermore one can assume (without loss of generality) 
that each of the
$U$'s, $V$'s and $W$'s has a definite total degree. Note that all
these degrees are nonnegative because the $U$'s, $V$'s and $W$'s
do not involve antifields and because it is assumed that
antighosts and Nakanishi--Lautrup fields have been 
eliminated already.

Again, the $(U^\ell, V^\ell)$ are called trivial pairs.
In order to deal with weak equalities the following
lemma will be useful later on:

\medskip
{\bf Lemma 5.2:} {\em Any weakly vanishing local 
total form $f(U,V,W)$ is
a combination of
weakly vanishing functions $L_K(W)$ in the sense that
\beq f(U,V,W)\approx 0\quad\Leftrightarrow\quad 
f(U,V,W)=a^K(U,V,W)L_K(W),
\quad L_K(W)\approx 0
\label{iii}\eeq
for some local total forms $a^K$.}

{\bf Proof:}
Since the classical equations of motion have vanishing
total degree and do not involve antifields, 
they are expressible solely
in terms of the
$U$'s and $W$'s because the $V$'s have positive total degrees
as a direct consequence of (\ref{iia})
(in fact only those $U$'s and $W$'s
with vanishing total degrees can occur in the equations of motion).
To prove (\ref{iii}) it is therefore sufficient to consider
functions depending only on the $U$'s and $W$'s. Now, if a
function $f(U,W)$ vanishes weakly
then the same holds for its $\4\gamma$-transformation due to 
the second identity in (\ref{cms}), for the latter implies
$f=\delta g\Rightarrow \4\gamma f=-\delta (\4\gamma g)\approx 0$.
Using (\ref{iia}) and (\ref{iib}) one concludes
\beq f(U,W)\approx 0 \quad\Rightarrow\quad \4\gamma f(U,W)=
V^\ell\frac{\6f(U,W)}{\6U^\ell}+R^i(W)\frac{\6f(U,W)}{\6W^i}
\approx 0.
\label{iiia}\eeq
Since the $U$'s, $V$'s and $W$'s are by assumption independent
local jet coordinates, and since the $V$'s do not
occur in the equations of motion, one concludes from
(\ref{iiia}) (for instance by differentiating
$\4\gamma f(U,W)$ w.r.t. to $V^\ell$)
that $f(U,W)\approx 0$ implies ${\6f(U,W)}/{\6U^\ell} \approx 0$.
Iteration of the argument yields
\beq f(U,W)\approx 0 \quad\Rightarrow\quad
\frac{\6^k f(U,W)}{\6U^{\ell_1}\ldots \6U^{\ell_k}} \approx 0
\quad\forall\, k\ .\label{iiid}\eeq
Thus a weakly vanishing function $f(U,W)$ must be a combination
of weakly vanishing functions of the $W$'s which
proves (\ref{iii}).\hspace*{\fill} $\Box$
\medskip

I remark that lemma 5.2 implies in particular that the equations
of motion themselves are equivalent to a set of
equations involving only those $W$'s
with vanishing total degree. 
This result will be interpreted in section
\ref{EOM} as the covariance of the equations
of motion. We are now prepared to prove that the $U$'s and $V$'s
can be eliminated from the weak $\4\gamma$-cohomology:
\medskip

\noindent {\bf Lemma 5.3:} {\em Suppose there is a
local and locally
invertible change of jet coordinates replacing
$\{[\Phi^A],x^\mu,dx^\mu\}$ by a set
$\{U^\ell, V^\ell, W^i\}$ satisfying (\ref{iia}) and (\ref{iib}).
Then locally the $U$'s and $V$'s can be eliminated from the
weak $\4\gamma$-cohomology on antifield 
independent local total forms,
\beq
\4\gamma \4\omega_0(U,V,W)\approx 0\quad\Rightarrow\quad
\4\omega_0(U,V,W)\approx f(W)+\4\gamma\4\eta_0(U,V,W),
\eeq
i.e.\ locally this cohomology is represented by solutions of}
\beq \4\gamma f(W)\approx 0,\quad
f(W)\not\approx \4\gamma g(W)+constant.\label{cov0}\eeq

{\bf Proof:} By assumption, locally any
antifield independent local total
form can be written in terms of the $U$'s, $V$'s and $W$'s.
To construct a contracting homotopy a parameter $t$ is introduced
scaling the $U$'s and $V$'s according to
\beq U^\ell_t:=tU^\ell,\quad V^\ell_t:=tV^\ell\ .\eeq
On total forms $\4\omega_0(U_t,V_t,W)$ 
one then defines an operator $b$
through
\beq b=U^\ell \frac{\6}{\6V^\ell_t}=
\frac 1t\, U^\ell \frac{\6}{\6V^\ell}\ .\label{bdef}\eeq
$\4\gamma U^\ell_t$ and $\4\gamma V^\ell_t$ 
are defined by replacing
in $\4\gamma U^\ell$ and $\4\gamma V^\ell$
all $U$'s and $V$'s by the corresponding $U_t$'s and
$V_t$'s. Now, (\ref{iia}) implies
$\4\gamma V^\ell=\4\gamma^2 U^\ell\approx 0$.
Using lemma 5.2 one thus concludes
\[ \4\gamma V^\ell=a^{\ell,K}(U,V,W)L_K(W),\quad
L_K(W)\approx 0 \]
for some $a^{\ell,K}$ and $L_K$. Hence one defines
\[\4\gamma U^\ell_t= V^\ell_t,\quad
\4\gamma V^\ell_t=a^{\ell,K}(U_t,V_t,W)L_K(W).\]
This shows in particular $\4\gamma V^\ell_t\approx 0$ and
one now easily verifies
\beq (\4\gamma b+b\4\gamma)\, \4\omega_0(U_t,V_t,W)\approx
\frac{\6\4\omega_0(U_t,V_t,W)}{\6 t}
\label{sb}\eeq
which implies
\beq \4\omega_0(U,V,W)-\4\omega_0(0,0,W)
\approx\int_0^1dt\, (\4\gamma b+b\4\gamma) \,  
\4\omega_0(U_t,V_t,W).
\label{reconst}\eeq
Applying again lemma 5.2 one concludes
that $\4\gamma \4\omega_0(U,V,W)\approx 0$
implies $\4\gamma \4\omega_0(U_t,V_t,W)\approx 0$. Using this
in (\ref{reconst}) we finally get
\beq \4\gamma \4\omega_0(U,V,W)\approx 0\quad\Rightarrow\quad
\4\omega_0(U,V,W)\approx
\4\omega_0(0,0,W)+\4\gamma\int_0^1dt\, b\, \4\omega_0(U_t,V_t,W)
\eeq
where we used
$\int_0^1dt\, \4\gamma b\4\omega_0(\ldots)\approx
\4\gamma\int_0^1dt\, b \4\omega_0(\ldots)$ (the latter
holds since $\4\gamma$ does not change the $t$-dependence
up to weakly vanishing terms). This proves the lemma.
\hspace*{\fill} $\Box$
\medskip

\noindent Remarks:

a) It is very important to realize that both
(\ref{IIA}) and (\ref{IIB}) must hold in order to 
eliminate $\cU$'s and $\cV$'s from the cohomology, and that the
existence of a pair of jet coordinates satisfying (\ref{IIA})
does in general {\em not} guarantee the existence of
complementary $\cW$'s fulfilling (\ref{IIB}).
A simple and important counterexample is given by
$x^\mu$ and $dx^\mu$ which always satisfy $\4s x^\mu =dx^\mu$
but usually do not form a trivial pair except in
diffeomorphism invariant theories, cf.\ section \ref{xdep}.
Analogous remarks apply of course to (\ref{iia}) and (\ref{iib}).
The reader may check that the contracting homotopies
for $s$ used in \cite{hental,ten,dragon} are in fact also based on
the construction of variables satisfying requirements
analogous to (\ref{iia}) and (\ref{iib}).

b) Clearly the aim is the construction of a set of local jet
coordinates containing as many trivial pairs as possible.
The difficulty of this construction is in general {\em not} the
finding of pairs of local jet coordinates satisfying (\ref{IIA})
resp.\ (\ref{iia})
but the construction of complementary $\cW$'s resp.\ $W$'s
satisfying (\ref{IIB}) resp.\ (\ref{iib}). 

c) Typically the
$U$'s are components of  gauge fields and their
derivatives and the $V$'s contain the corresponding derivatives
of the ghosts, cf.\ section \ref{ex}.
The $W$'s will be interpreted as tensor fields and
generalized connections, cf.\ section \ref{ten}.

d) Lemmas 5.1 and 5.3 are not always devoid of global subtleties,
i.e.\ they can fail to be globally valid.
E.g.\ if the manifold of the $U$'s
has a nontrivial de Rham cohomology, one cannot always
eliminate all the $U$'s and $V$'s globally (important
counterexamples are the vielbein fields in gravitational
theories, cf.\ \cite{bbhgrav}, section 5).
In such cases the proof of lemma 5.3 breaks down globally because
some of the functions of the $U$'s, $V$'s and $W$'s occurring in
the proof have no globally well-defined extensions.
This problem can be dealt with along the lines of
\cite{bbhgrav}.

\mysection{Gauge covariant algebra, tensor fields and generalized
connections}\label{ten}

It will now be shown
that the existence of a set of local jet coordinates
$\{U^\ell,V^\ell,W^i\}$ (with nonempty subset $\{U^\ell,V^\ell\}$)
satisfying (\ref{iia}) and (\ref{iib})
has a deep origin.
Namely it is intimately related to an algebraic
structure encoded in (\ref{iib}) which will
be interpreted as a gauge covariant algebra and leads to the
identification of tensor fields and generalized connections
mentioned in the introduction. 

Recall that each local jet coordinate $W^i$ has a
definite nonnegative total degree since it neither
involves antifields nor antighosts.
Those $W$'s with vanishing total degree
are called {\em tensor fields} and are denoted by $\cT^{\TIX}$;
the other $W$'s are called {\em generalized connections} for
reasons which will become clear soon. Those
generalized connections with total degree 1 are denoted by
$\4C^N$; the other
generalized connections are denoted by $\4Q^{N_G}$
where $G$ indicates their total degree,
\bea \{\cT^\TIX\}&=&\{W^i:\ \tot(W^i)=0\},\nonumber\\
\{\4C^N\}&=&\{W^i:\ \tot(W^i)=1\},\nonumber\\
\{\4Q^{N_G}\}&=&\{W^i:\ \tot(W^i)=G\geq 2\}.
\label{cov1}\eea
Note that the tensor fields have necessarily 
vanishing ghost number and
form degree, whereas a generalized connection decomposes
in general into a sum of local forms with different 
ghost numbers and
corresponding  form degrees,
\bea \4C^N&=&\7C^N+\cA^N,\quad
\gh(\7C^N)=1,\quad \gh(\cA^N)=0,\label{cov2a}\\
\4Q^{N_G}&=&\sum_{p=0}^G \7Q^{N_G}_p,\quad \gh(\7Q^{N_G}_p)=G-p.
\label{cov2b}\eea
The $\7C^N$ are called {\em covariant ghosts}, the $\cA^N$
{\em connection 1-forms} and the $\7Q^{N_G}_G$
{\em connection $G$-forms}.

Since $\4\gamma$ raises the total degree by one unit,
(\ref{iib}) and (\ref{cov1}) imply in particular
\bea \4\gamma\, \cT^\TIX&=&\4C^{N}\cR_{N}{}^\TIX(\cT),
\label{cov3}\\
\4\gamma\, \4C^{N}&=&\s0 12(-)^{\varepsilon_L+1}\4C^{L}\4C^{K}
\cF_{KL}{}^{N}(\cT)+\4Q^{M_2}\cZ_{M_2}{}^{N}(\cT),
\label{cov4}\\
\4\gamma\, \4Q^{N_2}&=&
\s0 12 (-)^{\varepsilon_L+1}\4C^K\4C^L\4C^M\cZ_{MLK}{}^{N_2}(\cT) 
\nonumber\\
& &+\4Q^{M_3}\cZ_{M_3}{}^{N_2}(\cT)                         
+\4Q^{M_2}\4C^K\cZ_{KM_2}{}^{N_2}(\cT)
\label{cov4a}\\
&\vdots&\nonumber
\eea
for some functions $\cR$, $\cF$ and $\cZ$ of the tensor fields.
Here $(\varepsilon_M+1)$ denotes the Grassmann parity of $\4C^M$,
\beq \varepsilon(\4C^M)=\varepsilon_M+1\ .\label{grading}\eeq
{}From $\4\gamma^2\cT^\TIX\approx 0$ one
concludes, using (\ref{cov3}) and (\ref{cov4}),
\bea & &\cR_{M}{}^\TJX\frac{\6\cR_{N}{}^\TIX}{\6\cT^\TJX}
-(-)^{\varepsilon_M\varepsilon_N}
\cR_{N}{}^\TJX\frac{\6\cR_{M}{}^\TIX}{\6\cT^\TJX}
\approx -\cF_{MN}{}^{K}\cR_{K}{}^\TIX\ ,\label{cov5}\\
& &\cZ_{M_2}{}^N\cR_N{}^\TIX\approx 0.
\label{cov5a}\eea
(\ref{cov5}) can be written in the compact form
\beq [\Delta_{M},\Delta_{N}]\approx -\cF_{MN}{}^{K}(\cT)\Delta_{K}
\label{cov6}\eeq
where $[\cdot,\cdot]$ denotes the graded commutator,
\beq [\Delta_{M},\Delta_{N}]=
\Delta_M\Delta_N-(-)^{\varepsilon_M\varepsilon_N}
\Delta_N\Delta_M\ ,
\label{cov6a}\eeq
and $\Delta_N$ is the operator
\beq \Delta_N=\cR_N{}^\TIX(\cT)\frac{\6}{\6\cT^\TIX}\ .
\label{cov7}\eeq
Analogously $\4\gamma^2\4C^N\approx 0$ implies in particular
\beq \csum {MNP}{5}
\left(\Delta_M\cF_{NP}{}^K+\cF_{MN}{}^R\cF_{RP}{}^K
+\cZ_{MNP}{}^{M_2}\cZ_{M_2}{}^K\right)\approx 0
\label{jacobi}\eeq
where the graded cyclic sum was used defined by
\beq\csum {MNP}{5}X_{MNP}=(-)^{\varepsilon_M\varepsilon_P}X_{MNP}+
(-)^{\varepsilon_N\varepsilon_M}X_{NPM}
+(-)^{\varepsilon_P\varepsilon_N}X_{PMN}\ .
\label{cyclic}\eeq
(\ref{jacobi}) are nothing but the Jacobi identities for the
algebra (\ref{cov6}) in presence of possible reducibility
relations (\ref{cov5a}).
Note that the Grassmann parities of $\4\gamma$ and
of the $\4C$'s imply the following
Grassmann parities and symmetries of the $\Delta$'s and $\cF$'s
\bea & & \varepsilon(\Delta_N)=\varepsilon_N\ ,\quad
\varepsilon(\cF_{MN}{}^K)=\varepsilon_M+\varepsilon_N
+\varepsilon_K\ \quad (mod\ 2),
\nonumber\\
& & \cF_{MN}{}^K=-(-)^{\varepsilon_M\varepsilon_N}\cF_{NM}{}^K\ .
\label{grading2}\eea

In order to reveal the geometric content of this
algebra it is useful to decompose (\ref{cov3}) and (\ref{cov4})
into parts with definite ghost numbers. Note that
(\ref{cov3}) reads
\beq \4\gamma\, \cT^\TIX=\4C^N \Delta_N \cT^\TIX \label{cov8a}\eeq
and thus decomposes due to $\4\gamma=\gamma+d$ 
and (\ref{cov2a}) into
\bea \gamma\, \cT^\TIX&=&\7C^N \Delta_N \cT^\TIX\ ,\label{cov8}\\
d\, \cT^\TIX&=&\cA^N \Delta_N \cT^\TIX\ .\label{cov9}\eea
(\ref{cov8}) can be interpreted as
a characterization of tensor fields as {\em gauge
covariant quantities}. Indeed, recall that tensor fields
are constructed solely out of the `classical fields' $\phi$, their
derivatives and the spacetime coordinates
due to (\ref{cov1}).
Therefore $\gamma \cT$ equals just a gauge transformation
of $\cT$ with parameters replaced by ghosts. (\ref{cov8})
requires thus that the gauge transformation of a tensor
field involves only special
combinations of the parameters and their derivatives (which may
involve the classical fields too),
corresponding to the covariant
ghosts $\7C$. 
Hence, (\ref{cov8}) characterizes tensor fields
indeed through a specific transformation law.

Now, the derivatives $\6_\mu\cT$ of a tensor field
are in general not tensor fields since $\gamma(\6_\mu\cT)$
contains $\6_\mu\7C^N$. The question arises how to
relate $\6_\mu\cT$ to gauge covariant quantities.
The answer is encoded in (\ref{cov9}).
Indeed, recall that the $\cA^N$ are 1-forms,
\beq \cA^N=dx^\mu \AA N\mu\ .\label{cov10}\eeq
(\ref{cov9}) is therefore equivalent to
\beq \6_\mu \cT^\TIX=\AA N\mu \Delta_N \cT^\TIX\ .\label{cov11}\eeq
By assumption, (\ref{cov11}) holds  {\em identically} in
the fields and their derivatives, with the same
set $\{\AA N\mu\}$ for all $\TIX$. 
In general this requires that
$\{\AA N\mu\}$ contains a locally invertible subset $\{\e m\mu\}$.
Then (\ref{cov11}) just defines those $\Delta$'s
corresponding to $\{\e m\mu\}$
in terms of the $\6_\mu$ and the other $\Delta$'s,
and can be regarded as a definition of {\em covariant derivatives}.
To put this in concrete terms I introduce the notation
\beq \{\AA N\mu\}=\{\e m\mu,\A \Indx\mu\},\quad
\{\Delta_N\}=\{\cD_m,\Delta_\Indx\},\quad m=1,\ldots,D
\label{cov12}\eeq
where the matrix
$(\e m\mu)$ is assumed to be invertible.
The $\cD_m$ are called covariant derivatives and according to
(\ref{cov11}) they are given by
\beq \cD_m=\E m\mu
  (\6_\mu-\A \Indx\mu\Delta_\Indx)\label{cov11a}\eeq
where $\E m\mu$ denotes the inverse of $\e m\mu$,
\beq 
\e m\mu\E m\nu=\delta_\mu^\nu\ ,\quad
\e m\mu\E n\mu=\delta_n^m\ .
\label{cov13}\eeq
I note that neither
the $\e m\mu$ nor the $\A \Indx\mu$ 
are necessarily elementary fields.
In particular, some of them may be constant or even zero.

Let me finally discuss (\ref{cov4}) which 
generalizes the Russian formula (\ref{i1}). Its decomposition
into pieces with definite ghost number (resp.\ form degree)
reads
\bea
& &\gamma \7C^{N}=\s0 12(-)^{\varepsilon_L+1}\7C^{L}\7C^{K}
\cF_{KL}{}^{N}(\cT)+\7Q^{M_2}\cZ_{M_2}{}^{N}(\cT),
\label{cov14}\\
& &\gamma \AA N\mu= \6_\mu\7C^N-\7C^{L}\AA K\mu
\cF_{KL}{}^{N}(\cT)-\7C_{\mu}{}^{M_2}\cZ_{M_2}{}^{N}(\cT),
\label{cov15}\\
& &\6_\mu \AA N\nu-\6_\nu \AA N\mu=-\AA L\mu \AA K\nu
\cF_{KL}{}^{N}(\cT)+B_{\mu\nu}{}^{M_2}\cZ_{M_2}{}^{N}(\cT)
\label{cov16}\eea
where the following notation was used:
\beq \4Q^{N_2}=\s0 12dx^\mu dx^\nu B_{\mu\nu}{}^{N_2}
+dx^\mu \7C_{\mu}{}^{N_2}+\7Q^{N_2}\ .
\label{cov17}\eeq
(\ref{cov14}) and (\ref{cov15}) give the $\gamma$-transformations
of the covariant ghosts and of the $\AA N\mu$
respectively.
(\ref{cov16}) determines the
{\em curvatures} (field strengths) corresponding to
the ``gauge fields" $\AA N\mu$. They are given by
\bea\cF_{mn}{}^N&=&\E m\mu\E n\nu\left(
2\6_{[\mu} \AA N{\nu]}
+2\e k{[\mu} \A \Indx{\nu]}\cF_{\Indx k}{}^{N}(\cT)\right.
\nonumber\\
& &\left.+\A \Indx\mu \A \Jndx\nu\cF_{\Jndx\Indx}{}^{N}(\cT)
-B_{\mu\nu}{}^{M_2}\cZ_{M_2}{}^{N}(\cT)\right)
\label{cov18}\eea
where the invertibility of the $\e m\mu$ was used again in order
to solve (\ref{cov16}) for the $\cF_{mn}{}^N$. That the latter
should indeed be identified with curvatures follows from the
fact that they occur in the commutator of the 
covariant derivatives,
\beq [\cD_m,\cD_n]\approx -\cF_{mn}{}^N\Delta_N\ .\label{cov19}\eeq
Note however that some (or all) of these curvatures may be constant
or even zero.
The Bianchi identities arising from (\ref{cov19}) 
are a subset of the identities (\ref{jacobi}),
\beq \cD_{[m}\cF_{nk]}{}^N-\cF_{[mn}{}^M \cF_{k]M}{}^N+
\cZ_{mnk}{}^{M_2}\cZ_{M_2}{}^N\approx 0.\label{bianchi}\eeq

\noindent Remarks:

a) (\ref{cov6}) can be regarded as a covariant version
of the gauge algebra. However it is important to realize that
the number of $\Delta$'s exceeds in general the number of
gauge symmetries, cf.\ section \ref{ex}.

b) $\4Q$'s occur only in reducible gauge theories
because otherwise there are no local jet variables which can
correspond to them.

c) Considerations similar to those performed here for the
$W$'s can be of course also
applied to the $\cW$'s satisfying (\ref{IIB}).
That leads in particular to an extension
of the concept to antifield dependent
tensor fields. Examples can be found
in \cite{rapid,bbhgrav,paper1}.

\mysection{Examples}\label{ex}

The concept outlined in the previous sections will now
be illustrated for four examples, exhibiting different
facets of the general formalism.
First the concept is  shown to reproduce
the standard tensor calculus in the familiar cases of Yang--Mills
theory and of gravity in the metric formulation.
Then pure four dimensional N=1 supergravity
without auxiliary fields is discussed. This illustrates the case
of an open gauge algebra and is the only
example where the number of $\Delta$'s and 
gauge symmetries coincide. Finally
Weyl and diffeomorphism invariant sigma models in two spacetime
dimensions are considered. In this example one gets
an infinite set
of generalized connections and corresponding 
$\Delta$-transformations,
but no (nonvanishing) curvatures (\ref{cov18}). I remark that
the approach of \cite{ten} does not apply to
any of these examples (not even to
gravity in the metric formulation!) because each of them violates
one of the assumptions (a)--(c) mentioned in section \ref{other}.
Hence, one really needs the extended concept outlined in the
previous sections to perform the following analysis.

As the gauge algebra is closed
in the first, second and last example, the
formulae of section \ref{ten} are in these cases promoted
to strict instead of weak equalities, with $\4\gamma$ replaced
by $\4s$ and without making reference to a 
particular gauge invariant
action.

\subsection{Yang--Mills theories}\label{ym}

For simplicity I consider pure Yang--Mills theories 
(no matter fields).
The standard BRST transformations of the
Yang--Mills gauge fields $\A i\mu$ and
the corresponding ghosts $C^i$ read
\beq s\A i\mu=\6_\mu C^i+C^k\A j\mu\f jki,\quad
sC^i=\s0 12\, C^kC^j\f jki
\label{ym1}\eeq
where $i$ labels the elements of the Lie algebra of the
gauge group with structure constants $\f ijk$.
The trivial pairs are in this case given by
\bea \{U^\ell\}&=&\{\6_{(\mu_1\ldots\mu_k}\A i{\mu_{k+1})}:
\ k=0,1,\ldots\},
\label{ym2}\\
\{V^\ell\}&=&\{\4s U^\ell\}=
\{ \6_{\mu_1\ldots\mu_{k+1}}C^i+\ldots:\ k=0,1,\ldots\}.
\label{ym3}\eea
Hence, in the new set of local jet coordinates the $V$'s replace
one by one all the derivatives of the ghosts. The undifferentiated
ghosts themselves are replaced by the generalized connections
\beq \4C^i=C^i+A^i,\quad A^i=dx^\mu\A i\mu\ .\label{ym4a}\eeq
The complete set of generalized connections contains 
in addition the
differentials,
\beq \{\4C^N\}=\{dx^\mu,\4C^i\}.\label{ym4}\eeq
The $\e m\mu$ are thus in this case just the entries of
the constant unit matrix, $\e m\mu=\delta_\mu^m$.
Hence, indices $m$ and $\mu$ need not be distinguished
in this case. The $\Delta$-operations 
corresponding to (\ref{ym4}) are
\beq \{\Delta_N\}=\{\cD_\mu,\delta_i\},\quad
\cD_\mu=\6_\mu-\A i\mu\delta_i\label{ym7}\eeq
where the $\delta_i$ are the Lie algebra elements.
(\ref{cov4}) reproduces for $N=i$
the ``Russian formula"  (\ref{i1})
in the form
\beq \4s\4C^i
=\s0 12\, \4C^k\4C^j\f jki
+\s0 12\, dx^\mu dx^\nu F_{\mu\nu}{}^i\ .\label{russian}\eeq
The algebra (\ref{cov6}) of the $\Delta$'s reads in this case
\beq [\cD_\mu,\cD_\nu]=-F_{\mu\nu}{}^i\delta_i,\quad
[\cD_\mu,\delta_i]=0,\quad [\delta_i,\delta_j]=\f ijk\delta_k
\label{ym8}\eeq
with the standard Yang--Mills field strengths
arising from (\ref{cov18}) and transforming 
under the $\delta_i$ according
to the adjoint representation,
\beq F_{\mu\nu}{}^i=
\6_\mu\A i\nu-\6_\nu\A i\mu+\A j\mu\A k\nu\f jki\ ,\quad
\delta_i F_{\mu\nu}{}^j=-\f ikj F_{\mu\nu}{}^k\ .
\label{ym9}\eeq
A complete set of tensor fields is in this case given by
the $x^\mu$ and a choice of algebraically independent components
of the field strengths and their covariant
derivatives,
\beq \{\cT^\TIX\}\subset\{x^\mu,
\cD_{\mu_1}\ldots\cD_{\mu_k}F_{\nu\rho}{}^i:\ k=0,1,\ldots\}.
\label{ym10}\eeq

\noindent Remark:

Notice that the above choice of variables is very similar to the
one in \cite{hental}. In fact the tensor fields
coincide in both approaches (except that here also the $x^\mu$ 
are counted among them). The difference is that the present approach
singles out the $\4C^i$ and $dx^\mu$ as generalized
connections, rather than just the $C^i$. Note that, as a 
direct consequence of the presence of $d$ in $\4s$,
one {\em cannot} simply choose $\4C^i=C^i$ here because that
choice would not fulfill requirement (\ref{iib}).

\subsection{Gravity in the metric formulation}\label{metgrav}

 I consider now pure gravity with
the metric fields $g_{\mu\nu}=g_{\nu\mu}$
as the only classical fields and diffeomorphisms as the only
gauge symmetries.
The BRST transformations of the metric and the diffeomorphism
ghosts $\xi^\mu$ read
\beq sg_{\mu\nu}=\xi^\rho\6_\rho g_{\mu\nu}+
(\6_\mu \xi^\rho)\, g_{\rho\nu}+(\6_\nu \xi^\rho)\, g_{\mu\rho}\ ,
\quad s\xi^\mu=\xi^\nu\6_\nu\xi^\mu\ .\label{met1}\eeq
The trivial pairs can be chosen as
\bea \{U^\ell \}&=&\{x^\mu,\6_{(\mu_1\ldots\mu_k}
\chris {\mu_{k+1}}{\mu_{k+2})}\nu:\ k=0,1,\ldots\}\label{met2}\\
\{V^\ell\}&=&\{\4s U^\ell\}=
\{dx^\mu, \6_{\mu_1\ldots\mu_{k+2}}\xi^\nu+\ldots\ :\ 
k=0,1,\ldots\}
\label{met3}\eea
where
\beq
\chris \mu\nu\rho=\s0 12\, g^{\rho\sigma}
(\6_\mu g_{\nu\sigma}+\6_\nu g_{\mu\sigma}-\6_\sigma g_{\mu\nu}).
\label{met4}\eeq
Note that the $V$'s replace all derivatives of the ghosts
of order $>1$. The undifferentiated ghosts and their first order
derivatives give rise to the generalized connections
\beq \{ \4C^N\}=\{\4\xi^\mu,\4C_\mu{}^\nu\},\quad
\4\xi^\mu=\xi^\mu+dx^\mu,\quad
\4C_\mu{}^\nu=\6_\mu \xi^\nu+\chris \mu\rho\nu\4\xi^\rho\ .
\label{met5}\eeq
The generalized Russian formulae (\ref{cov4}) read in this case
\beq \4s\4\xi^\mu=\4\xi^\nu \4C_\nu{}^\mu,\quad
\4s\4C_\mu{}^\nu=\4C_\mu{}^\rho\4C_\rho{}^\nu+
\s0 12 \4\xi^\rho\4\xi^\sigma R_{\rho\sigma\mu}{}^\nu
\label{met13}\eeq
where $R_{\mu\nu\rho}{}^\sigma$ is the standard Riemann tensor
constructed of the $\Gamma$'s.
The $\e m\mu$ are, as in the case of the Yang--Mills theory, just
the entries of the constant unit matrix. Hence,
indices $\mu$ and $m$ are not distinguished. One gets
\beq \{\Delta_N\}=\{\cD_\mu,\Delta_\mu{}^\nu\},\quad
\cD_\mu=\6_\mu-\chris \mu\rho\nu \Delta_\nu{}^\rho
\label{met8}\eeq
where the $\Delta_\mu{}^\nu$ generate
$GL(D)$-transformations  of world
indices according to
\beq \Delta_\mu{}^\nu T_\rho=\delta_\rho^\nu T_\mu\ ,\quad
\Delta_\mu{}^\nu T^\rho=-\delta_\mu^\rho T^\nu\ .\label{met7}\eeq
The algebra (\ref{cov6}) reads now
\bea & & [\cD_\mu,\cD_\nu]=
         -R_{\mu\nu\rho}{}^\sigma \Delta_\sigma{}^\rho\ ,
\nonumber\\
& & [\Delta_\mu{}^\nu,\cD_\rho]=\delta_\rho^\nu \cD_\mu\ ,\quad
[\Delta_\mu{}^\rho,\Delta_\nu{}^\sigma]=
\delta_\nu^\rho \Delta_\mu{}^\sigma-
\delta_\mu^\sigma \Delta_\nu{}^\rho
\ .\label{met9}\eea
The set of tensor fields contains the $g_{\mu\nu}$, $\mu\geq\nu$
and a maximal
set of algebraically independent components of
$R_{\mu\nu\rho}{}^\sigma$ and their covariant derivatives,
\beq \{\cT^\TIX\}\subset\{g_{\mu\nu},
\cD_{\mu_1}\ldots \cD_{\mu_k}
R_{\lambda\nu\rho}{}^\sigma:\ k=0,1,\ldots\}.
\label{met12}\eeq

\noindent Remark:

Recall that tensor fields are
characterized by the transformation law (\ref{cov8}). One might
wonder whether this transformation law agrees in this case
with the standard transformation law for tensor fields under
diffeomorphisms which is in BRST language the Lie derivative
along the diffeomorphism ghosts.
The answer is affirmative because
(\ref{cov8}) yields in this case, e.g.\ for a tensor field $T_\mu$,
\beq \gamma\, T_\mu= \xi^\nu\cD_\nu T_\mu+
(\6_\mu \xi^\nu+\chris \mu\rho\nu\xi^\rho)T_\nu
=\xi^\nu\6_\nu T_\mu+(\6_\mu\xi^\nu)T_\nu .
\eeq

\subsection{D=4, N=1 minimal supergravity}\label{sugra}

The classical field content of the D=4, N=1 
minimal pure supergravity theory without
auxiliary fields is given by
the vielbein fields and the gravitinos, denoted by $\viel a\mu$
and $\psi_\mu{}^\alpha$, $\5\psi_\mu{}^{\dot{\alpha}}$ respectively
($\alpha, \dot{\alpha}$ denote indices
of two-component complex
Weyl spinors with conventions as
in \cite{SUGRA}).
The gauge symmetries are diffeomorphism invariance,
local supersymmetry and local Lorentz
invariance. The corresponding ghosts
are denoted by $\xi^\mu$, $\xi^\alpha$, $\5\xi^{\dot{\alpha}}$
and $C^{ab}=-C^{ba}$
respectively. For simplicity the analysis is
restricted to the action \cite{sugraold}
\beq S_{cl}=\int d^4x\, \left[  \s0 12\, e R
-2\epsilon^{\mu\nu\rho\sigma}
(\psi_\mu\sigma_\nu \nabla_\rho\5\psi_\sigma
-\5\psi_\mu\5\sigma_\nu \nabla_\rho\psi_\sigma)\right]
\label{sugra1}\eeq
with $e=\det (\viel a\mu)$, $\epsilon^{0123}=1$ and
\bea
& &R=R_{ab}{}^{ba}\ ,\quad
R_{ab}{}^{cd}=2\Viel  {[a}\mu\Viel {b]}\nu
(\6_\mu{\omega_\nu}^{cd} +{\omega_\mu}^{ce}{\omega_{\nu e}}^d),
\label{sugra3a}\\
& &\nabla_\mu\psi_\nu{}^\alpha=\6_\mu \psi_\nu{}^\alpha-
\s0 12\, \omega_\mu{}^{ab}
\sigma_{ab\, \beta}{}^\alpha \psi_\nu{}^\beta,
\label{sugra3b}\\
& &\nabla_\mu\5\psi_\nu{}^{\dot \alpha}=
\6_\mu \5\psi_\nu{}^{\dot \alpha}+
\s0 12\, \omega_\mu{}^{ab}
\5\sigma_{ab}{}^{\dot \alpha}{}_{\dot \beta}
\5\psi_\nu{}^{\dot \beta}
\label{sugra3c}
\eea
where the
$\Viel a\mu$ are the entries of the inverse vielbein and
$\omega_\mu{}^{ab}$ denotes the
gravitino dependent spin connection
\bea \omega_\mu{}^{ab}&=&
E^{a\nu}E^{b\rho}(\omega_{[\mu\nu]\rho}-\omega_{[\nu\rho]\mu}+
\omega_{[\rho\mu]\nu}),\nonumber\\
\omega_{[\mu\nu]\rho}&=& e_{\rho a}\6_{[\mu} \viel a{\nu]}
-i\psi_\mu\sigma_\rho\5\psi_\nu+i\psi_\nu\sigma_\rho\5\psi_\mu\ .
\label{sugra2}\eea
(Lorentz indices $a,b,\ldots$ are lowered and raised with the
Minkowski metric $\eta_{ab}=diag(1,-1,-1,-1)$.)

The $\gamma$-transformations read in this case
\bea
\gamma\viel a\mu&=&\xi^\nu\6_\nu\viel a\mu
              +(\6_\mu\xi^\nu)\viel a\nu+
              C_b{}^a\viel b\mu+2i\sigma^a{}_{\alpha{\dot \alpha}}
              (\xi^\alpha\5\psi_\mu{}^{\dot \alpha}
              -\5\xi^{\dot \alpha}\psi_\mu{}^\alpha)\ ,
                                             \label{sugra4}\\
\gamma\psi_\mu{}^\alpha&=& \nabla_\mu\xi^\alpha+
                     \xi^\nu\6_\nu\psi_\mu{}^\alpha
                     +(\6_\mu\xi^\nu)\psi_\nu{}^\alpha
  +\s0 12\, C^{ab}\sigma_{ab\, \beta}{}^\alpha \psi_\mu{}^\beta\ ,
                                                   \label{sugra5}\\
\gamma\xi^\mu&=&\xi^\nu\6_\nu\xi^\mu
           +2i\xi^\alpha\sigma^\mu{}_{\alpha{\dot \alpha}}
           \5\xi^{\dot \alpha}\ ,
                                                   \label{sugra6}\\
\gamma\xi^\alpha&=&\xi^\mu\6_\mu \xi^\alpha
           +\s0 12\, C^{ab}\sigma_{ab\beta}{}^\alpha \xi^\beta
           -2i\xi^\beta\sigma^\mu{}_{\beta {\dot \beta}}
           \5\xi^{\dot \beta}
           \psi_\mu{}^\alpha\ ,
                                                   \label{sugra7}\\
\gamma C^{ab}&=&\xi^\mu\6_\mu C^{ab}-C^{ac}C_c{}^b
           -2i\xi^\beta\sigma^\mu{}_{\beta{\dot \beta}}
           \5\xi^{\dot \beta}
           \omega_\mu{}^{ab}
\label{sugra8}\eea
(and analogous expressions for $\gamma\5\psi_\mu$ and
$\gamma\5\xi$) where
\[ \nabla_\mu\xi^\alpha=\6_\mu \xi^\alpha-
\s0 12\, \omega_\mu{}^{ab}\sigma_{ab\, \beta}{}^\alpha \xi^\beta\ .
\]
The gauge algebra is open (it closes modulo the equations of motion
for the gravitinos). Hence $\gamma$ is nilpotent only on-shell
and does not agree with $s$ on all the fields.

One can choose the $U$'s in this case as
\bea \{U^\ell\}&=&\{x^\mu,\,
\6_{(\mu_1\ldots\mu_{k}}\viel a{\mu_{k+1})},\,
\6_{(\mu_1\ldots\mu_k}\omega_{\mu_{k+1})}{}^{cd},\  \nonumber\\
& &\phantom{\{}
\6_{(\mu_1\ldots\mu_k}\psi_{\mu_{k+1})}{}^\alpha,\,
\6_{(\mu_1\ldots\mu_k}\5\psi_{\mu_{k+1})}{}^{\dot \alpha}:
\ c>d;\ k=0,1,\ldots \}.
\label{grav3}\eea
Note that the $\omega_\mu{}^{ab}=\omega_\mu{}^{[ab]}$
correspond one by one
to the antisymmetrized first order derivatives
$\6_{[\mu}\viel a{\nu]}$ of the vielbein fields due to 
(\ref{sugra2}).
Hence, all the
$U^\ell$ are indeed algebraically independent new local
jet coordinates. The corresponding $V^\ell$ replace one by one
the $dx^\mu$ and all the derivatives of the ghosts due to
$\4\gamma\viel a\mu=\6_\mu\xi^a+\ldots$ 
($\xi^a=\viel a\mu\xi^\mu$),
$\4\gamma\omega_\mu{}^{ab}=\6_\mu C^{ab}+\ldots$\ ,
$\4\gamma\psi_\mu{}^\alpha= \6_\mu\xi^\alpha+\ldots$
and $\4\gamma\5\psi_\mu{}^{\dot \alpha}=
-\6_\mu\5\xi^{\dot \alpha}+\ldots$\ .
The undifferentiated ghosts give rise to the generalized
connections
\bea
& &\{\4C^N\}=\{\4\xi^a,\, \4\xi^\alpha,\, \4\xi^{\dot \alpha},\,
\4C^{ab}:\ a>b\},
\nonumber\\
& &\4\xi^a=\4\xi^\mu\viel a\mu\ ,\quad
\4C^{ab}=C^{ab}+\4\xi^\mu\omega_\mu{}^{ab}\ ,\nonumber\\
& &\4\xi^\alpha=\xi^\alpha+\4\xi^\mu\psi_\mu{}^\alpha\ ,\quad
\4\xi^{\dot \alpha}=\5\xi^{\dot \alpha}-
\4\xi^\mu\5\psi_\mu{}^{\dot \alpha}
                                              \label{sugra9}\eea
with $\4\xi^\mu$ as in (\ref{met5}). The corresponding $\Delta$'s
are denoted by
\bea 
& &\{\Delta_N\}=\{\cD_a,\, \cD_\alpha,\, \5\cD_{\dot \alpha},\,
l_{ab}:\ a>b\},
                                            \nonumber\\
& &\cD_a=\Viel  a\mu(\6_\mu-\s0 12\omega_\mu{}^{ab}l_{ab}
-\psi_\mu{}^\alpha\cD_\alpha+
\5\psi_\mu{}^{\dot \alpha}\5\cD_{\dot \alpha})
                                              \label{sugra10}\eea
where $l_{ab}=-l_{ba}$ denote the elements of the Lorentz algebra,
and $\cD_\alpha$ and $\5\cD_{\dot \alpha}$ are
supersymmetry transformations represented on the
tensor fields given below (these tensor fields are ordinary fields,
not superfields; accordingly $\cD_\alpha$ and $\5\cD_{\dot \alpha}$
are not `superspace operators'). The Grassmann parities
of the $\Delta$'s are $\varepsilon_a=\varepsilon_{[ab]}=0$ and
$\varepsilon_\alpha=\varepsilon_{\dot \alpha}=1$
(the supersymmetry ghosts commute).
(\ref{sugra10})
indicates that in this case the vielbein fields are identified
with the $\e m\mu$, i.e.\ the indices $m$ 
coincide here with Lorentz
vector indices,
\beq\e m\mu\equiv \viel a\mu\ ,\quad \E m\mu\equiv \Viel a\mu\ .
\label{grav7a}\eeq
Using the shorthand notation
\[ \{\4\xi^A\}=\{\4\xi^a,\4\xi^\alpha,\4\xi^{\dot \alpha}\},\quad
\{\cD_A\}=\{\cD_a,\cD_\alpha,\5\cD_{\dot \alpha}\}\]
the algebra of the $\cD_A$ reads
\beq  [\cD_A,\cD_B]\approx
-T_{AB}{}^C\cD_C-\s0 12F_{AB}{}^{cd}l_{cd}\label{sugra11}\eeq
where the nonvanishing $T_{AB}{}^C$ and $F_{AB}{}^{cd}$ are
\bea
T_{\alpha{\dot \alpha}}{}^a &=&T_{{\dot \alpha}\alpha}{}^a=
2i\sigma^a{}_{\alpha{\dot \alpha}}\ ,\\
T_{ab}{}^\alpha&=&\Viel a\mu\Viel b\nu
       (\nabla_\mu\psi_\nu{}^\alpha-\nabla_\nu\psi_\mu{}^\alpha),
                                                 \label{sugra12a}\\
F_{{\dot \alpha} b}{}^{cd}&=&i\, (T^{cd\alpha}
       \sigma_{b\, \alpha{\dot \alpha}}
       -2\sigma^{[c}{}_{\alpha{\dot \alpha}} T^{d]}{}_b{}^\alpha) ,
                                                  \\
F_{ab}{}^{cd}&=&
  R_{ab}{}^{cd}+2(\psi_{[a}{}^\alpha F_{b]\alpha}{}^{cd}
  -\5\psi_{[a}{}^{\dot \alpha} F_{b]{\dot \alpha}}{}^{cd})
                                              \label{sugra12}\eea
and analogous expressions for $T_{ab}{}^{\dot \alpha}$ and
$F_{\alpha b}{}^{cd}$. The remaining commutators
of the $\Delta$'s are
\beq [l_{ab},\cD_A]=-g_{[ab]\, A}{}^B\cD_B\ ,\quad
[l_{ab},l_{cd}]=2\eta_{a[c}l_{d]b}-2\eta_{b[c}l_{d]a}
                                             \label{sugra13a}\eeq
where
\beq g_{[ab]\, c}{}^d=2\eta_{c[a}\delta^d_{b]}\ ,\quad
g_{[ab]\, \alpha}{}^\beta=\sigma_{ab\, \alpha}{}^\beta\ ,\quad
g_{[ab]\, {\dot \alpha}}{}^{\dot \beta}=
-\5\sigma_{ab}{}^{\dot \beta}{}_{\dot \alpha} \ .
                                             \label{sugra13b}\eeq
Accordingly the generalized Russian formulae (\ref{cov4})
read in this case
\bea \4\gamma\, \4\xi^A  &=&\s0 12\4C^{ab}g_{[ab]\, B}{}^A \4\xi^B
-\s0 12(-)^{\varepsilon_B} \4\xi^B \4\xi^CT_{ CB}{}^A\ ,
                                                \label{sugra14}\\
\4\gamma\, \4C^{ab}  &=&-\4C^{ac}\4C_c{}^b
         -\s0 12(-)^{\varepsilon_D}\4\xi^D\4\xi^C F_{CD}{}^{ab}\ .
                                              \label{sugra15}\eea
Note that these identities
encode all the equations (\ref{sugra2})--(\ref{sugra8}),
(\ref{sugra12a}) and (\ref{sugra12}).
The set of independent tensor fields 
consists in this case of a subset
of $F_{ab}{}^{cd}$, $T_{ab}{}^\alpha$,
$T_{ab}{}^{\dot \alpha}$ and their covariant derivatives,
\bea \{\cT^\TIX\}&\subset&
\{\cD_{a_1}\ldots \cD_{a_k} F_{bc}{}^{de},\,
\cD_{a_1}\ldots \cD_{a_k} T_{bc}{}^\alpha\ ,\nonumber\\
& &\ \cD_{a_1}\ldots \cD_{a_k} T_{bc}{}^{\dot \alpha}:
\ k=0,1,\ldots\}.\label{sugra16}\eea
\medskip

\noindent Remark:

Notice that the formalism provides `super-covariant' tensor fields
and, in particular, `super-covariant' derivatives (\ref{sugra10}) 
containing the gravitino and the supersymmetry
transformations. 
Note also that these tensor fields do not carry 
``world indices" $\mu,\nu,\ldots$\ , in contrast to
the example discussed in the previous subsection.
The reason is that the undifferentiated vielbein fields
count among the $U$'s. Indeed, the corresponding $V$'s replace
all the first order derivatives of the diffeomorphism ghosts
$\xi^\mu$ and therefore the BRST transformation of a tensor field
must not involve $\6_\nu \xi^\mu$. Hence,
tensor fields are indeed `world scalars' in this case. One 
could of course
instead count the undifferentiated vielbein fields also among the
tensor fields and promote the $\6_\nu \xi^\mu$ to generalized
connections.
Then tensor fields could also carry world indices and 
one would get additional $\Delta$'s generating
$GL(4)$ transformations of world indices,
as in the metric formulation of gravity discussed in the
previous subsection. However, such a choice would not correspond
to a maximal set of trivial pairs
and would thus complicate unnecessarily the
analysis of the BRST cohomology!

\subsection{Two dimensional sigma models}\label{sigma}

Consider two dimensional sigma models whose set of classical fields
consists of scalar fields $\varphi^i$ and the two dimensional
metric fields $g_{\mu\nu}$ and whose gauge symmetries are given by
two dimensional diffeomorphism and Weyl invariance,
with corresponding ghosts $\xi^\mu$ and $C$ respectively.
The BRST transformations of the fields read
\bea & &sg_{\mu\nu}=\xi^\rho\6_\rho g_{\mu\nu}+
(\6_\mu \xi^\rho)\, g_{\rho\nu}+(\6_\nu \xi^\rho)\, g_{\mu\rho}
+C\, g_{\mu\nu}\ ,\nonumber\\
& &sY=\xi^\nu\6_\nu Y\quad \mbox{for}\quad 
Y\in\{\varphi^i,\xi^\mu,C\}.
\label{sigma1}\eea
Following closely the lines (but not the notation) of
\cite{paper1} I first introduce
new local jet coordinates $h,\5h,e,\eta,\5\eta$
replacing the undifferentiated metric components
and diffeomorphism ghosts
($h,\5h$ are ``Beltrami variables")%
\footnote{This change of jet coordinates is
not globally well-defined in general.},
\bea & & h=\frac{g_{11}}{g_{12}+\sqrt {g}}\ ,\quad
\5h=\frac{g_{22}}{g_{12}+\sqrt{g}}\ ,\quad e=\sqrt{g},
\label{sigma2a}\\
& & \eta=(\xi^1+dx^1)+\5h(\xi^2+dx^2),\quad
\5\eta=(\xi^2+dx^2)+h(\xi^1+dx^1)\label{sigma2b}\eea
with $g=-\det (g_{\mu\nu})>0$. The $U$'s are
\beq \{U^\ell\}=\{x^\mu,\, \6^p\5\6^qh,\,
\6^p\5\6^q\5h,\, \6^p\5\6^qe:\ p,q=0,1,\ldots\}
\label{sigma3}\eeq
where
\beq \6\equiv \6_1,\quad \5\6\equiv \6_2\ .\label{not}\eeq
Hence, in this case
all the metric components and all their derivatives
occur in trivial pairs. The corresponding
$V$'s replace one by one the
$C,\6\5\eta,\5\6\eta$, all their derivatives, and the $dx^\mu$.
Therefore one gets in this example an infinite set of
generalized connections, given by $\eta$, $\5\eta$ and their
remaining derivatives,
\bea & & \{\4C^N\}=\{\eta^p,\5\eta^{\5 p}:\ 
p,{\5 p}=-1,0,1,\ldots\},
\label{sigma4}\\
&& \eta^p=\s0 1{(p+1)!}\, \6^{p+1}\eta,\quad
\5\eta^{\5 p}=\s0 1{({\5 p}+1)!}\, \5\6^{{\5 p}+1}\5\eta\ .
\label{sigma5}\eea
(\ref{sigma1})--(\ref{sigma2b}) imply
$\4s\eta=\eta\6\eta$ and
$\4s\5\eta=\5\eta\5\6\5\eta$. Therefore
(\ref{cov4}) reads in this case
\beq \4s\eta^p=\frac 12\sum_{r=-1}^{p+1}(p-2r)\eta^r\eta^{p-r}
\label{sigmax}\eeq
and an analogous formula for $\4s\5\eta^{\5 p}$.
The infinite set of $\Delta$'s corresponding to the $\4C$'s 
is denoted by
\beq \{\Delta_N\}=\{L_p,\5L_{\5 p}:\ p,{\5 p}=-1,0,1,\ldots\}.
\label{sigma6}\eeq
Recall that the r.h.s. of (\ref{sigmax}) contains the
structure functions occurring in the algebra of the $\Delta$'s. 
In this
case {\em all} of these functions are constant and
the algebra of the $L$'s and $\5L$'s is isomorphic 
to two copies of the
algebra of regular vector fields $(-z^{p+1})\6/\6z$,
\beq [L_p,L_q]=(p-q)L_{p+q}\ ,\quad
[\5L_{\5 p},\5L_{\5q}]=({\5 p}-\5q)\5L_{{\5 p}+\5q}\ ,\quad
[L_p,\5L_{\5 p}]=0.\label{sigma7}\eeq
The set of tensor fields on which this algebra is realized
is given by
\beq \{\cT^\TIX\}=\{T^i_{p,{\5 p}}:\ p,{\5 p}=0,1,\ldots\},\quad
T^i_{p,{\5 p}}=(L_{-1})^p (\5L_{{-1}})^{\5 p}\varphi^i\ .
\label{sigma8}\eeq
The explicit form of the $T^i_{p,{\5 p}}$ in terms of the
fields and their derivatives was discussed in \cite{paper1}
and will be rederived below for the first few $T$'s.
The algebraic representation of the $L$'s and $\5L$'s on the tensor
fields can be derived from the algebra (\ref{sigma8}) using
$L_p T^i_{0,0}=\5L_{\5 p} T^i_{0,0}=0$ $\forall\, p,{\5 p}\geq 0$.
The latter follows from the identification $\4s T^i_{0,0}=
\4C^N \Delta_N T^i_{0,0}$, cf.\ (\ref{cov8a}). This yields
\beq q<p:\quad  L_qT^i_{p,{\5 p}}=
\frac {p!}{(p-q-1)!}\, T^i_{p-q,{\5 p}}\ ; \quad
q\geq p:\quad L_qT^i_{p,{\5 p}}=0
\label{sigma10a}\eeq
and analogous formulae for  $\5L_{\5q}T^i_{p,{\5 p}}$.

Let us now make contact with section \ref{ten}.
Using (\ref{sigma2b}) one easily reads off from (\ref{sigma5})
the connection forms $\cA^p$ and $\5\cA^{\5 p}$
contained in $\eta^p$ and $\5\eta^{\5 p}$:
\bea & &\cA^p=\delta_{-1}^p\, dx^1+H^p\, dx^2 ,\quad
\5\cA^{\5 p}=\delta_{{-1}}^{\5 p}\, dx^2+\5H^{\5 p}\, dx^1\ ,
\label{sigmaconn1}\\
& & H^p=\s0 1{(p+1)!}\, \6^{p+1}\5h,\quad
\5H^{\5 p}= \s0 1{({\5 p}+1)!}\, \5\6^{{\5 p}+1}h.
\label{sigmaconn2}\eea
The components of $\cA^{-1}$ and $\5\cA^{{-1}}$
are identified with the $\e m\mu$ according to
\beq \{\e m\mu\}\equiv\{\cA^{-1}_\mu,{\5\cA_\mu}^{{-1}}\}.
\label{sigmaconn3}\eeq
Explicitly one thus gets in matrix form
\beq (\e m\mu)=\left(\ba{cc} 1 & h \\ \5h & 1  \ea\right),\quad
(\E m\mu)=\frac{1}{1-h\5h}\left(\ba{cc} 1 & -h \\ -\5h & 1  
\ea\right).
\label{sigmaconn4}\eeq
Due to (\ref{sigmaconn3}) the covariant derivatives 
$\{\cD_m\}\equiv
\{\cD,\5\cD\}$ are identified with
$L_{-1}$ and $\5L_{-1}$. (\ref{cov11a}) yields now
\beq \cD=\frac{1}{1-h\5h}\left(
\6-h\5\6-\sum_{{\5 p}\geq 0}\5H^{\5 p} \5L_{\5 p}
+h\sum_{p\geq 0}H^p L_p \right)\label{sigmacov2}\eeq
and an analogous expression for $\5\cD$, with
$H$'s as in (\ref{sigmaconn2}). Thanks to (\ref{sigma10a}), the
occurrence of infinitely many $L$'s and $\5L$'s
in $\cD$ and $\5\cD$ does not result in nonlocal
expressions (tensor fields). Note that
$\cD\equiv L_{-1}$ and $\5\cD\equiv \5L_{-1}$ commute 
according to (\ref{sigma7}), i.e.\ in this case all
the (infinitely many) curvatures (\ref{cov18}) vanish!
Due to (\ref{sigma8}) the set of tensor fields is given by
the $\varphi^i$ and all their covariant derivatives. The
latter may now be constructed
recursively using (\ref{sigma10a}).
With $V=(1-h\5h)^{-1}$ one gets for instance (derivatives
act on everything to the right):
\bea
T^i_{1,0}&=&\cD\varphi^i=V(\6-h\5\6)\varphi^i\ ,\quad
T^i_{0,1}=\5\cD\varphi^i=V(\5\6-\5h\6)\varphi^i\nonumber\\
T^i_{1,1}&=&\5\cD\cD\varphi^i=V(\5\6-\6\5h)V(\6-h\5\6)\varphi^i
\nonumber\\
&=&\cD\5\cD\varphi^i=V(\6-\5\6 h)V(\5\6-\5h \6)\varphi^i\ .
\label{sigma10}\eea

\mysection{Structure of the solutions}
\label{structure}

In this section some
conclusions are drawn about the geometric structure
of the solutions of the cohomological problem
and the related physical quantities such as
gauge invariant actions, conserved currents
and anomalies.

According to section \ref{red} the cohomological
problem in question can be reduced to the solution of
$\4\gamma\4\omega_0\approx 0$ where $\4\omega_0$ does not
depend on antifields. This cocycle condition decomposes
into  the ``weak descent equations"
\beq d\alpha_{M}\approx 0,\quad 
\gamma\alpha_{p}+d\alpha_{p-1}\approx 0\quad \mbox{for}
\ p_0<p\leq M,\quad
\gamma\alpha_{p_0}\approx 0
\label{str0}\eeq
where $\alpha_{p}$ denotes the $p$-form contained in $\4\omega_0$,
\beq \4\omega_0=\sum_{p=p_0}^M \alpha_{p}\ ,\quad
M=\min \{D,G\},\quad G=\tot(\4\omega_0).\label{str1}\eeq
Here $M=\min \{D,G\}$ holds because
$\4\omega_0$ does not contain antifields\footnote{I also assume,
without loss of generality, that $\4\omega_0$ does not
depend on antighosts or Nakanishi--Lautrup fields 
(cf.\ section \ref{trivpairs}) and
that it has a definite total degree.}
and thus involves only
$p$-forms with $p\leq G$.
Note that the first equation in (\ref{str0}) is trivially
satisfied if $M=D$, i.e.\ if $G\geq D$.

Now, the only local forms which are
weakly $d$-closed but not necessarily (locally)
weakly $d$-exact are volume forms and
forms which do not depend on the ghosts, i.e.
all other weakly $d$-closed local forms are
also weakly $d$-exact,
\beq d\alpha_{q}\approx 0,\quad q<D,\quad \gh(\alpha_{q})>0
\quad\Rightarrow\quad \alpha_{q}\approx d\eta_{q-1}\ .
\label{hen}\eeq
This follows by means of the algebraic Poincar\'e lemma
(cf.\ section \ref{descent}) immediately from
a general result on the relative cohomology of $\delta$ and $d$
derived in \cite{hen91}.
Since all the local forms $\alpha_{p}$ with
$p<M$ occurring in (\ref{str1})
have positive ghost number,
one can analyse the weak descent
equations (\ref{str0}) by means of (\ref{hen}) like
the usual descent equations by means of the algebraic
Poincar\'e lemma in section \ref{descent}.
This leads to the conclusion
that $\4\omega_0$ is a nontrivial solution of
$\4\gamma\4\omega_0\approx 0$ if and only if its part
$\alpha_{M}$ fullfills
\bea G> D:& & \gamma\alpha_{D}+d\alpha_{D-1}\approx 0,\quad
\alpha_{D}\not\approx \gamma\eta_{D}+d\eta_{D-1}\ ;
\label{str2}\\
G= D:& & \gamma\alpha_{D}+d\alpha_{D-1}\approx 0,\quad
\alpha_{D}\not\approx d\eta_{D-1}\ ;
\label{act1}\\
G<D:& & d\alpha_{G}\approx 0,\quad
\alpha_{G}\not\approx d\eta_{G-1}+constant.
\label{str3}\eea
Furthermore one concludes, using (\ref{hen}) again, that
all solutions of (\ref{str2})--(\ref{str3}) can be
completed to nontrivial
solutions of $\4\gamma\4\omega_0\approx 0$. Hence,
the complete local BRST cohomology is in fact (locally)
isomorphic to the cohomological problems
established by (\ref{str2})--(\ref{str3}).

Let me now discuss the implications of sections \ref{trivpairs} and
\ref{ten} for the structure of the solutions of
(\ref{str2})--(\ref{str3}) and briefly comment on their
physical interpretation.
For notational convenience I will restrict this discussion to
the case of an irreducible gauge algebra. Using the notation
of section \ref{ten}, one can
then assume $\4\omega_0$ to be of the form
\beq\4\omega_0=
\4C^{N_1}\ldots \4C^{N_G}a_{N_G\ldots N_1}(\cT),\quad
\4C^N=\7C^N+\cA^N.
\label{str4}\eeq
This implies that, in irreducible gauge theories,  
the general solutions of (\ref{str2})--(\ref{str3}) are  
of the form
\bea G> D: & & \alpha_{D}=
\cA^{N_1}\ldots \cA^{N_D}\7C^{N_{D+1}}\ldots \7C^{N_G}
a_{N_G\ldots N_1}(\cT);\label{str5}\\
G= D: & & \alpha_{D}=
\cA^{N_1}\ldots \cA^{N_D}
a_{N_D\ldots N_1}(\cT);\label{act2}\\
G< D: & & \alpha_{G}=
\cA^{N_1}\ldots \cA^{N_G} a_{N_G\ldots N_1}(\cT),
\label{str6}
\eea
up to trivial and ``topological" 
(= locally but not globally trivial) solutions, of course.
(\ref{str5}) applies for
$G=D+1$ to the antifield independent part of
integrands of candidate gauge anomalies. Well-known examples are 
representatives of chiral
anomalies in Yang--Mills theory
\cite{StoraZumino}. Their integrands have indeed 
the form (\ref{str5}) 
(recall that in Yang--Mills theory the differentials
count among the connection forms, cf.\ subsection \ref{ym}).

Solutions of (\ref{act1}) give rise to
BRST invariant functionals with ghost number 0 and
thus (\ref{act2}) applies to
integrands of gauge invariant actions and 
their continuous first order
deformations \cite{bh}. However,
concerning these solutions a few more remarks
are in order which I postpone to the next section.

The solutions of (\ref{str3}) provide the local
conservation laws of the theory.
They correspond for $G=D-1$ one-to-one to the nontrivial
conserved currents%
\footnote{A conserved current $j^\mu$ ($\6_\mu j^\mu\approx 0$)
is called trivial in this context if
$j^\mu\approx \6_\nu S^{\nu\mu}$ holds for some local
$S^{\nu\mu}=-S^{\mu\nu}$.} and generalize for smaller $G$
the concept of nontrivial conserved currents to form
degrees $<D-1$ \cite{bbh1}.

We conclude that all nontrivial ``dynamical" conserved
local $G$-forms can be written in the form (\ref{str6})
if the gauge algebra is irreducible, and in a similar form, 
involving
possibly connection forms of higher form degree,
if the gauge algebra is reducible (``topological'' conserved
local forms cannot always be cast
in this form). In fact, in ``normal'' theories,
dynamical solutions of (\ref{str3})
exist at most at form degrees $G\geq D-(2+r)$ where $r$ denotes the
reducibility order of the theory, see \cite{bbh1}
($r=-1$ for theories without
gauge invariance, $r=0$ for irreducible gauge theories, \ldots).
The weak $d$-cohomology established by (\ref{str3}) goes
sometimes under the name ``characteristic cohomology" 
\cite{charac}.

To illustrate the result on the conservation laws I consider the
Noether current corresponding to the invariance of the
supergravity action (\ref{sugra1}) under global 
$U(1)$-transformations
of the gravitino. One finds for this Noether current
$j^\mu$ and the corresponding solution $\alpha_{3}$ of (\ref{str3})
\bea j^\mu&=& -2i\epsilon^{\mu\nu\rho\lambda}
              \psi_\nu\sigma_\rho\5\psi_\lambda\ ,\label{str9}\\
\alpha_{3}
 &=& \s0 1{6}\, dx^\mu dx^\nu dx^\rho
     \epsilon_{\mu\nu\rho\lambda}\, j^\lambda
    =2i e^a\psi\sigma_a\5\psi
\label{str10}\eea
where $\epsilon^{0123}=-\epsilon_{0123}=1$ and
\beq e^a=dx^\mu \viel a\mu,\quad \psi^\alpha
=dx^\mu\psi_\mu{}^\alpha,\quad
\5\psi^{\dot \alpha}=-dx^\mu\5\psi_\mu{}^{\dot \alpha}\ .
\label{str11}\eeq
$\alpha_{3}$ is indeed of the form (\ref{str6}) because the
1-forms (\ref{str11}) are among the connection forms $\cA^N$
of the supergravity theory, cf.\ subsection \ref{sugra}. Note that
this solution of (\ref{str3}) is constructed solely out of
connection forms, i.e.\ it does not involve tensor fields at all!

\noindent Remark:

The usual construction of Noether
currents does not always provide the corresponding
solutions of (\ref{str3})
directly in the form (\ref{str6}). The statement here is that
one can always redefine the Noether
currents by subtracting trivial currents (if necessary)
such that the corresponding $(D-1)$-forms
take the geometric form (\ref{str6}). A famous example
for such a redefined current is the ``improved" energy momentum
tensor in Yang--Mills theory.

\mysection{Structure of gauge invariant actions}
\label{action}

The field-antifield formalism is usually constructed starting
from a given gauge invariant classical action. 
One may then ask whether
it is possible to deform this action without destroying
the gauge invariance. This question is relevant
for instance in the quantum theory where deformations of the
action can be caused by quantum corrections, or for the deformation
of free gauge theories to interacting ones.
The BRST cohomology provides a powerful tool to tackle
these problems \cite{bh}. 

One may distinguish two kinds of
deformations of a given action:
those which do not change the gauge transformations
up to local field redefinitions, and
those which modify simultaneously the action and the gauge
transformations in a nontrivial way. 

The integrands (volume forms) of
actions which are invariant under {\em given} 
gauge transformations have to satisfy
\beq \gamma\alpha_{D}+d\alpha_{D-1}= 0,\quad
\alpha_D\neq d\eta_{D-1}\ ,\quad
\alpha_{D}=d^D\! x\, a(x,[\phi])
\label{act3}\eeq
where $\gamma$ encodes the gauge transformations under study.
Note that (\ref{act3}) is a stronger
condition than (\ref{act1}) and replaces the latter for two
reasons: the integrands of gauge invariant
actions are (i) required to be {\em strictly}
$\gamma$-invariant up to a total derivative and
(ii) not necessarily to be considered
as trivial if they are weakly zero up to a total derivative --
for instance one would not call the Einstein--Hilbert
action $\int d^4x \sqrt{-g}R$ trivial even though its integrand
is weakly zero. 

Now, if the gauge algebra is (off-shell) {\em closed}, $\gamma$ is
strictly nilpotent on all the fields (but not necessarily on the
antifields). Therefore (\ref{act3}) implies descent equations
for $\gamma$ and $d$ which do not involve antifields and
read in a compact form $\4\gamma\4\omega_0=0$,
$\4\omega_0\neq\4\gamma\4\eta_0$. This problem can be analysed
like the weak $\4\gamma$-cohomology
in sections \ref{trivpairs} and \ref{ten} -- all the
arguments go through also for
strict instead of weak equalities since $\gamma$ and $\4\gamma$
are strictly nilpotent on all the fields. In particular
we conclude that the general solution of (\ref{act3}) has again the
form (\ref{act2}) (up to $d$-exact contributions)
if the gauge algebra is closed and irreducible.
The general solution of (\ref{act3}) provides the most
general action which is invariant under a given set of
gauge transformations encoded in $\gamma$. It
has been determined by means of the
BRST cohomology for Yang--Mills theory \cite{com},
gravity \cite{grav}, minimal N=1, D=4 supergravity
\cite{sugra,SUGRA} (both in the old minimal formulation 
\cite{sugraaux} and in the new minimal one \cite{sugranew})
and for the sigma models considered in
subsection \ref{sigma} \cite{paper1}. One can check that
in all these cases the integrand of the most general
gauge invariant action can indeed be expressed in
the geometric form (\ref{act2}) even though this is not
completely obvious
in all cases. For instance, written in this form
the integrand of the supergravity action (\ref{sugra1}) reads
\bea & & \alpha_{4}=-\epsilon_{abcd}e^a e^b e^c
(\s0 1{48}e^d \cR
+\s0 i3 S\sigma^d\5\psi-\s0 i3\psi\sigma^d \5S),
\label{str13}\\
& & \cR=F_{ab}{}^{ba},\quad
S^\alpha=T_{ab}{}^\beta\sigma^{ab}{}_\beta{}^\alpha
\label{str13a}\eea
with $e^a$, $\psi$ and $\5\psi$ as in (\ref{str11}),
$T_{ab}{}^\alpha$ and $F_{ab}{}^{cd}$ as in (\ref{sugra12a})
and (\ref{sugra12}), and $\epsilon_{0123}=-1$.

The determination of
deformations of a given action which modify
nontrivially the gauge transformations
is more subtle. A method which allows to attack this problem
systematically and is based on the BRST cohomology
was outlined in \cite{bh}. The idea is to deform
the solution of the master equation instead of the classical
action itself. This has many advantages. In particular it
shows that to first order in the deformation parameter
the deformed action is required to be weakly invariant under the
original (undeformed) $\gamma$. The
integrand of this first order deformation thus
has the form (\ref{act2}) up to weakly vanishing terms.
However, in general an analogous statement does not apply to the
terms of higher orders in the deformation parameter because
these terms are not necessarily weakly $\gamma$-invariant.

\mysection{Gauge covariance of the equations of motion}\label{EOM}

A direct corollary  of lemma 5.2 is the gauge covariance of the
equations of motion
(cf.\ remark after the proof of that lemma).
Indeed, lemma 5.2 implies that
the classical equations of motion are equivalent to a set
of weakly vanishing functions of those $W$'s with vanishing
total degree. Since the latter are just the tensor fields
(cf.\ section \ref{ten}), we conclude:
\medskip

{\bf Lemma 10.1:} {\em The classical equations of motion in a
gauge theory are gauge covariant in the sense that they
are equivalent to a set of
weakly vanishing functions of the tensor fields.}
\medskip

This is of course well-known for standard gauge theories
such as Yang--Mills theory and
Einstein gravity where the Euler--Lagrange equations themselves
turn out to be expressible solely
in terms of the tensor fields.
A less trivial check of lemma 10.1 can be performed for the
supergravity action (\ref{sugra1}). Indeed one can verify
that the corresponding equations of motion are equivalent
to the following
equations involving only the tensor fields (\ref{sugra16}):
\beq \cR_{ab}\approx 0,\quad S^\alpha\approx 0,\quad
U_{{\dot \alpha}{\dot \beta}}{}^\alpha\approx 0
\label{str7}\eeq
with $S^\alpha$ as in (\ref{str13a}) and
\beq \cR_{ab}=F_{acb}{}^c,\quad
U_{{\dot \alpha}{\dot \beta}}{}^\alpha=
T_{ab}{}^\alpha\5\sigma^{ab}{}_{{\dot \alpha}{\dot \beta}}\ .
\label{str8}\eeq

\mysection{Discussion of $x$-dependence}\label{xdep}

This section is devoted to the discussion of
a special aspect of the cohomological problem
concerning the explicit dependence of the solutions
on the spacetime coordinates.
In particular it is emphasized that the result of the
cohomological analysis depends on whether it is
carried out in the space of $x$-dependent or
$x$-independent local forms. It is therefore important
to make clear in every analysis of the 
BRST cohomology in which space
one works and to be aware of the consequences of the chosen
approach.

\subsection{General remarks}

In general the results
of the cohomological analysis
will depend in two respects on whether or not one considers
the problem in the space of $x$-dependent local forms:
(a) some nontrivial representatives of the cohomology
might be overlooked if one performs the cohomological
analysis in the space of $x$-independent local forms;
(b) solutions which are nontrivial in the
space of $x$-independent local forms can become trivial
in the space of $x$-dependent local forms.

It is important to realize that (b) applies also
to theories which do not admit
solutions of the cohomological problem
depending nontrivially  on the $x^\mu$ at all.
An important subclass of theories
with this property
are those which are invariant under spacetime diffeomorphism.
They will
be discussed  in the next subsection in this context.

Let me first illustrate (a) and (b) for the simple
example of the free $D$-dimensional Maxwell action
\[ S_{maxwell}=\int d^D\! x F_{\mu\nu} F^{\mu\nu},\quad
F_{\mu\nu}=\6_\mu A_\nu-\6_\nu A_\mu\ .\]
Important examples for $x$-dependent solutions are the
Noether currents associated with the Lorentz invariance
of $S_{maxwell}$. In ``improved"  (= gauge
invariant) form these currents read
\[ j^\mu_{lorentz}=\lambda_{\rho\nu}x^\rho T^{\nu\mu},\quad
\lambda_{\rho\nu}=-\lambda_{\nu\rho}=constant \]
where
$ T^{\mu\nu}=F^{\rho\mu}{F_\rho}^\nu
-\s0 14\eta^{\mu\nu}F_{\rho\sigma} F^{\rho\sigma}$
is the ``improved" energy momentum tensor
($\eta^{\mu\nu}$ is the Minkowski metric).
The $x$-dependence of $j^\mu_{lorentz}$
cannot be removed by subtracting trivial currents from it.
This illustrates (a). To demonstrate (b) I consider
the $x$-independent current
\[ j^\mu_{triv}=\lambda_\nu F^{\nu\mu}, \quad 
\lambda_\nu=constant.\]
It is clearly
conserved too, $\6_\mu j^\mu_{triv}\approx 0$, and
the corresponding $(D-1)$-form is
nontrivial in the space of $x$-independent forms.
However it is trivial in the space of $x$-dependent forms, 
as one has
\[ j^\mu_{triv}\approx \6_\nu (\lambda_\rho x^\rho F^{\nu\mu}).\]
This reflects that the global symmetry of
$S_{maxwell}$ which corresponds via Noether's theorem to 
$j^\mu_{triv}$
is trivial too \cite{bbh1}: it is the
shift symmetry $A_\mu\rightarrow A_\mu+\lambda_\mu$ and
is trivial because it is
just a gauge transformation with parameter $\lambda_\mu x^\mu$.

\subsection{Implications of diffeomorphism invariance}\label{diff}

As mentioned already, in diffeomorphism invariant theories
(of the standard type)
one can remove any explicit $x$-dependence locally from all the
solutions of the cohomological problem by subtracting
trivial solutions. This was observed and used
first in \cite{grav} for the antifield independent BRST cohomology
in standard gravity. It is instructive to
see how this result arises naturally within the framework
of section \ref{trivpairs}. Namely
it follows simply from the fact that all the 
$x^\mu$ and $dx^\mu$ form
trivial pairs $(\cU^\ell,\cV^\ell)$
as a direct consequence of diffeomorphism invariance.
To see this, note first that $x^\mu$ and $dx^\mu$ indeed satisfy
requirement (\ref{IIA}),
\[ \4s\, x^\mu=dx^\mu\ .\]
Now, this is valid for any theory but does {\em not}
imply in general that $x^\mu$ and $dx^\mu$ form a trivial pair
because to that end (\ref{IIB}) must hold in as well.
However, in contrast to other theories 
(such as Yang--Mills theory),
one can usually fulfill
this additional requirement in diffeomorphism invariant theories
through a simple change of variables (jet coordinates):
one just replaces
the diffeomorphism ghosts $\xi^\mu$ with the combinations
\[ \4\xi^\mu=\xi^\mu+dx^\mu\ .\]
Indeed, in standard diffeomorphism invariant
theories the $\4s$-transformation of all the fields and
antifields\footnote{Assuming again that
antighosts and Nakanishi--Lautrup fields have been eliminated
from the cohomological problem already, 
cf.\ section \ref{trivpairs}.} depends
on the $\xi^\mu$ and $dx^\mu$ only via $\4\xi^\mu$ for one has
$sZ=\xi^\mu\6_\mu Z+\ldots$, $dZ=dx^\mu\6_\mu Z$ and thus
$\4sZ=\4\xi^\mu\6_\mu Z+\ldots$ for any field or antifield $Z$
(the nonwritten terms in $sZ$ do not contain undifferentiated
$\xi$'s in standard diffeomorphism invariant theories).
This reflects that the diffeomorphisms are encoded
in the BRST operator through the Lie derivative along
$\xi$ and implies that (i) $x^\mu$ and $dx^\mu$ indeed form
a trivial pair and can thus be eliminated locally from the
cohomology, i.e.\ the nontrivial solutions of $\4s\4\omega=0$
can be chosen so as not to depend explicitly on the $x^\mu$ and to
depend on $\xi^\mu$ and $dx^\mu$ only via $\4\xi^\mu$,
(ii) on $x$-independent functions and local total forms respectively,
$s$ and $\4s$ arise from each other through the replacements
$\xi^\mu\leftrightarrow \4\xi^\mu$, i.e.
\[ \4s=\rho\circ s\circ \rho^{-1},\quad
s=\rho^{-1}\circ \4s\circ \rho\]
where
\[ \rho=\exp \left(dx^\mu \frac{\6}{\6\xi^\mu}\right),\quad
\rho^{-1}=\exp \left(-dx^\mu \frac{\6}{\6\4\xi^\mu}\right).
\]
In particular this implies the now well-known result,
first derived in \cite{grav}, that the descent equations
go in standard diffeomorphism invariant theories
always down to a BRST-invariant $x$-independent 0-form $\omega_0$,
and that the ``integration" of
the descent equations starting from such a 0-form is not
obstructed and results in a solution $\4\omega=\rho \omega_0$
of  $\4s\4\omega=0$.

I stress however that this result is valid only
in the space of $x$-{\em dependent}
forms. Indeed, in the space of $x$-independent forms
there are {\em additional} solutions ``$\4\omega$ times monomial of
the $dx^\mu$" where $\4\omega$ is an $x$-independent solution
because in that space $dx^\mu$ is an ``$\4s$-singlet".
In particular it is not true
that the descent equations go always down to a 0-form
if one restricts the cohomological analysis
to $x$-independent forms.

\mysection{Conclusion}\label{conclusion}

The framework proposed in
this paper to analyse the local BRST cohomology 
is based on a few very simple ideas:
(i) the formulation of the local BRST cohomology in
the jet bundle approach, (ii) the mapping of
the BRST cohomology to the cohomology of $\4s=s+d$
and to its on-shell counterpart, the
antifield independent weak cohomology of $\4\gamma=\gamma+d$,
(iii) the construction of contracting
homotopies to eliminate certain
jet coordinates, called trivial pairs,
from the $\4s$-cohomology.

In spite of its conceptual simplicity, (iii) is not straightforward
because it requires the construction of an appropriate
set of local jet coordinates splitting into two subsets
one of which contains the trivial pairs whereas the other one 
consists of complementary jet coordinates which are required to
generate an $\4s$- resp.\ $\4\gamma$-invariant subalgebra and
are interpreted as tensor fields and generalized connections.
The existence (and finding) of such complementary
jet coordinates is a crucial prerequisite for the
elimination of trivial pairs and
was shown to be intimately related to a gauge covariant algebra.
The construction of such jet coordinates
has been illustrated for various examples to
demonstrate the proposed method and its large range of
applicability.

The outlined method
simplifies the computation of the BRST cohomology
considerably by reducing it locally to a cohomological problem
involving only the tensor fields and generalized connections.
The simplification does not only consist in the
fact that some jet coordinates, the trivial pairs, are eliminated.
On top of that, and equally important, one obtains
a very compact and useful formulation of the remaining
cohomological problem on tensor fields and generalized connections
through equations such as (\ref{cov3})--(\ref{cov4a}).
For specific models the compact formulation of the
BRST algebra obtained in this way is in fact well-known
in the literature.
For instance (\ref{cov4}) reproduces in the Yang--Mills case
the celebrated ``Russian formula" (\ref{i1}) 
which was used
especially within the algebraic construction
and classification of chiral
anomalies \cite{StoraZumino,talon}.

It should be remarked that this
simplifies the computation of the BRST cohomology, but of course
does not solve it. Nevertheless it allows
remarkable conclusions about the `geometric' structure and
covariance properties of the 
solutions of the cohomological problem and
the related physical quantities (Noether currents,
gauge invariant actions, candidate gauge anomalies,
etc.), as well as of the classical equations
of motion, cf.\ sections \ref{structure}--\ref{EOM}.

Finally I remark that local jet coordinates
with the mentioned properties are
also useful when one needs to take global (topological)
aspects into account which have been completely neglected in this
paper. In particular, 
global obstructions to the elimination of trivial
pairs may be taken into account using K\"unneth's 
theorem \`a la \cite{bbhgrav}.

\section*{Acknowledgement.}

This work was carried
out in the framework of the European Community Research Programme
``Gauge theories, applied supersymmetry and quantum gravity", 
with a
financial contribution under contract SC1-CT92-0789.

\appendix

\mysection{Fields, antifields and their jet space}\label{jet}

The so-called minimal set
of fields contains the `classical fields' $\phi^i$ which
occur in the gauge invariant classical
action, the ghosts $C^\alpha$ corresponding one-to-one to
the nontrivial gauge symmetries and the ghosts for ghosts
$Q^{\alpha_k}$ of first and higher order $k=1,\ldots,r$
where $r$ denotes the reducibility order of the
theory ($r=0$ for irreducible gauge theories),
\beq
\{\Phi^A\}_{\min}=
\{\phi^i,\, C^\alpha,\, Q^{\alpha_k} \},\quad
\{\Phi^*_A\}_{\min}=\{\phi^*_i,\, C^*_\alpha,\, Q^*_{\alpha_k}\}.
\label{jet1}\eeq
In order to fix the gauge one usually extends the minimal
set of fields to a nonminimal one by adding
antighosts and Nakanishi--Lautrup auxiliary fields.
Each field $\Phi^A$ has a definite Grassmann parity 
$\varepsilon(\Phi^A)$
and ghost number $\gh(\Phi^A)$.
The Grassmann parities and ghost numbers of the antifields are
related to those of the fields according to
\bea
& &\gh(\phi^i)=0,\quad \gh(C^\alpha)=1,\quad \gh(Q^{\alpha_k})=k+1,
\nonumber\\
& &\gh(\Phi^*_A)=-\gh(\Phi^A)-1,\quad
\varepsilon(\Phi^*_A)=\varepsilon(\Phi^A)+1\quad (mod\ 2).
\label{jet2}\eea
The Grassmann parity of the classical fields is 0 for bosonic
(commuting) fields and 1 for fermionic (anticommuting) fields,
the Grassmann parity of the ghosts is opposite to the Grassmann
parity of the corresponding gauge symmetry, and the
Grassmann parity of the ghosts for ghosts is determined
analogously. Fields and antifields commute or anticommute
according to their Grassmann parities,
\beq Z_1Z_2=(-)^{\varepsilon(Z_1)\varepsilon(Z_2)}Z_2Z_1\ .
\label{jet3}\eeq

The fields and antifields and all their derivatives are
considered as local coordinates of an
infinite jet space. For this set of
jet coordinates I use the
collective notation
\beq [\Phi,\Phi^*]\equiv\{
\6_{\mu_1\ldots\mu_k}\Phi^A,\,
\6_{\mu_1\ldots\mu_k}\Phi^*_A:\ k=0,1,\ldots\}
\label{des0}\eeq
and these jet coordinates are regarded as independent
apart from the identities
\beq \6_{\mu_1\ldots\mu_i\ldots\mu_j\ldots\mu_k}Z
=\6_{\mu_1\ldots\mu_j\ldots\mu_i\ldots\mu_k}Z\quad\forall\, i,j.
\label{des0b}\eeq
The derivatives $\6_\mu$ have vanishing Grassmann parity
and ghost number.
The set of local jet coordinates is completed
by the spacetime coordinates $x^\mu$ and
by the differentials $dx^\mu$ which are counted among the jet
coordinates by convenience. The former have even,
the latter odd Grassmann parity, both have
vanishing ghost number.

The derivatives $\6_\mu$ are
defined as total derivative operators in the jet space
according to
\beq \6_\mu=\frac{\6}{\6x^\mu}+\sum_{k\geq 0;\,
\nu_{i+1}\geq\nu_i}
(\6_{\mu\nu_1\ldots\nu_k}Z_I)\,
\frac{\6}{\6(\6_{\nu_1\ldots\nu_k}Z_I)}
\label{totder}\eeq
where $\{Z_I\}=\{\Phi^A,\Phi^*_A\}$. The sum
in (\ref{totder}) runs only over
those $\6_{\nu_1\ldots\nu_k}Z$ with $\nu_{i+1}\geq\nu_i$
because of the identities (\ref{des0b}).
It is further understood that
\beq \frac{\6(\6_{12}Z)}{\6(\6_{12}Z)}=
\frac{\6(\6_{21}Z)}{\6(\6_{12}Z)}=1\quad etc.\label{des0c}\eeq

\mysection{BRST operator}\label{brs}

The BRST operator is constructed from a solution
$S$ of the master equation \cite{bv} of the form
\beq S=\int d^D\! x\, \cL,\quad
\cL=\cL_{cl}(x,[\phi])-
(R_\alpha^i C^\alpha)\phi^*_i+\ldots \label{brs1}\\
\eeq
where $\cL_{cl}$ denotes the Lagrangian of the
gauge invariant classical action and
$R_\alpha^i C^\alpha$ is the gauge transformation of $\phi^i$
with gauge parameters replaced by the ghosts, i.e.\
$R_\alpha^i$ is an operator of the form
\beq R_\alpha^i(x,[\phi],\6) =\sum_{k\geq 0}
r_\alpha^{i\mu_1\ldots\mu_k}(x,[\phi])
\6_{\mu_1}\ldots\6_{\mu_k}\ .\label{brs3}\eeq
If the gauge algebra is reducible with
reducibility order $r$, then $\cL$ is required to
contain also a piece of the form
\beq
C^*_\alpha Z_{\alpha_1}^{\alpha} Q^{\alpha_1}+
\sum_{k=2}^r Q^*_{\alpha_{k-1}} Z_{\alpha_k}^{\alpha_{k-1}}
Q^{\alpha_k}, \label{brs4}\eeq
where the $Z$'s are operators
of the form (\ref{brs3})
implementing the reducibility relations.
For the purpose of gauge fixing one may also include
pieces involving antighosts and
Nakanishi--Lautrup auxiliary fields \cite{bv}.

The BRST transformations of $\Phi^A$ and $\Phi^*_A$ are given by
their antibrackets \cite{bv} with $S$ according to
$s\,\cdot=(S,\,\cdot\,)$. This results in
\beq s\, \Phi^A=-\frac {\7\6^R \cL}{\7\6 \Phi^*_A}\ ,\quad
s\, \Phi^*_A=\frac {\7\6^R \cL}{\7\6 \Phi^A}
\label{brs5}\eeq
where $\7\6^R\cL/\7\6 Z$ denotes the Euler--Lagrange 
right-derivative
of $\cL$ with respect to $Z$ (derivatives $\6^R/\6$ act 
from the right),
\beq \frac {\7\6^R \cL}{\7\6 Z}=\sum_{k\geq 0;\,
\mu_{i+1}\geq\mu_i}(-)^k\6_{\mu_1}\ldots\6_{\mu_k}
\frac {\6^R\cL}{\6(\6_{\mu_1\ldots\mu_k}Z)} \ .
\label{brs6}\eeq
The BRST transformations of derivatives of the 
fields and antifields
are obtained from (\ref{brs5}) simply by requiring 
$s\6_\mu=\6_\mu s$,
i.e.
\beq s(\6_\mu \Phi^A)=\6_\mu (s\Phi^A)=
-\6_\mu \, \frac {\7\6^R \cL}{\7\6 \Phi^*_A}\quad etc.
\label{brs7}\eeq

\end{document}